\documentclass[%
 reprint,
showpacs,%preprintnumbers,
 amsmath,amssymb,
 aps,
]{revtex4-1}

\usepackage{graphicx}% Include figure files
\usepackage{dcolumn}% Align table columns on decimal point
\usepackage{bm}% bold math
\newcommand{\be}{\begin{equation}}
\newcommand{\ee}{\end{equation}}
\newcommand{\bea}{\begin{eqnarray}}
\newcommand{\nn}{\nonumber}
\newcommand{\eea}{\end{eqnarray}}

\begin{document}

\preprint{APS/123-QED}

\title{Revising the multipole moments of numerical spacetimes, and its consequences}% Force line breaks with \\

\author{George Pappas}
\author{Theocharis A. Apostolatos}%
\affiliation{Section of Astrophysics, Astronomy, and Mechanics,
Department of Physics, University of Athens, Panepistimiopolis Zografos GR15783,
Athens, Greece}%

\date{\today}% It is always \today, today,

\begin{abstract}
Identifying the relativistic multipole moments
of a spacetime of an astrophysical object that has been constructed numerically
is of major interest, both because the multipole moments are
intimately related to the internal structure of the object, and
because the construction of a suitable analytic metric that mimics a numerical
metric should be based on the multipole moments of the
latter one, in order to yield a reliable representation.
In this note we show that there has been a widespread delusion
in the way the multipole moments
of a numerical metric are read from the asymptotic expansion of
the metric functions. We show how one should read
correctly the first few multipole moments (starting  from
the quadrupole mass-moment), and how these corrected moments improve the efficiency of
describing the metric functions with analytic metrics
that have already been used in the literature, as well as other consequences
of using the correct moments.
\end{abstract}

% 04.25.D-; 04.25.Nx; 04.40.Dg; 26.60.Kp}


\maketitle

%\address{Section of Astrophysics, Astronomy, and Mechanics,
%Department of Physics, University of Athens, Panepistimiopolis
%Zografos GR15783, Athens, Greece}

%\ead{gpappas@phys.uoa.gr}

%------------------------------------------------------------------------
\section{\label{sec_1}Introduction}

%------------------------------------------------------------------------

In the beginning of the 70s Geroch and Hansen defined the multipole moments of an asymptotically
flat spacetime in the static and stationary case in analogy to the newtonian ones
\cite{Gero,Hans}. In 1989 Fodor et al.~\cite{Fodoretal} found a
concise and practical way to compute the multipole moments of a spacetime that is additionally axially symmetric,
taking advantage of the insightful Ernst-potential formalism. The definitions of
alternative relativistic multipole moments
by Simon \cite{Simon} and by Thorne \cite{Thorne} should be noted as well;
the former one ends up to the same moments as
Geroch's and Hansen's, while Thorne's moments are coordinate dependent
and thus they could be used as a criterion to choose
the appropriate coordinates to read the usual Geroch-Hansen moments (for a review see \cite{Quevedo}).

The advent of  technological developments that offer us the capability to observe gravitational
wave signals gave a further boost to the study and use of the notion of relativistic
multipole moments during the last two decades.
The multipole moments uniquely characterize the gravitational
field of a compact object; thus Ryan \cite{Ryan} wrote formulae that
relate the moments of a spacetime with the observable
frequencies and the number of cycles of the
gravitational wave signal that is emitted by a low mass object inspiraling
adiabatically into such a spacetime. It should be emphasized that
both the moments and the frequencies in Ryan's paper are invariant
quantities that do not depend on the coordinates used to describe
the background metric. Besides Ryan's attempt to connect the
multipole moments with astrophysical observables, Shibata and Sasaki (SS) have given analytic relations
of the radius of the innermost stable circular orbit (ISCO) to the moments \cite{ShibSasa}, and Laarakkers and Poisson
have attempted to relate the multipole moments of a neutron star with the equation of state (EOS) of the matter
it consists of \cite{LaarPois}.

Exploring all these interconnections of multipole moments with a particular metric,
expressed either in an analytic form or through a numerical grid, brought into
our attention a systematic deviation of the way the multipole
moments are read from the asymptotic expansions of various
metrics. This systematic error arises mainly from erroneously assuming
that a metric, which is expressed in a given coordinate
system, has an asymptotic behavior similar to a Schwarzschild metric up to some order.

We have tried to correct such errors by relying on the
coordinate-invariant expressions of Ryan \cite{Ryan}. Thus one could obtain the correct moments by computing the
gravitational-wave spectrum $\Delta \tilde{E}$ (the energy emitted per unit logarithmic
frequency interval) of a test particle that is orbiting on a circular equatorial orbit in
an asymptotically flat, stationary and axially symmetric spacetime. Note that we do not assume that
the astrophysical object we study is actually
surrounded by such test particles emitting gravitational
radiation; we just use this hypothetical configuration to relate
quantities (frequencies and moments) which are
independent from the coordinates in which the metric is presented.

First we corrected the quadrupole-moment values of the rotating neutron star models, which were constructed by the
numerical code of Stergioulas \cite{SterFrie}. Then we showed that if one
tries to approximate the metric functions by a three-parameter
analytic metric, like the Manko et al. metric \cite{manko1},
the numerical metric is, in most cases, almost an order of magnitude better approximated by
that particular analytic metric than what was initially found in \cite{BertSter}.
This conclusion ensures that a suitable analytic metric with only a
few free parameters (three, or four \cite{inprep}) could be quite faithful to
represent with very good accuracy the gravitational field of a realistic neutron star.

The rest of the letter is organized as follows: %In Sec.~\ref{sec_2}
First we compare the asymptotic expressions for the metric functions, derived by Butterworth and Ipser (BI)
\cite{ButtIpse}, to the corresponding asymptotic expressions introduced by Komatsu, Eriguchi and Hechisu  (KEH) \cite{KEH}.
The KEH formalism was later implemented numerically by Cook, Shapiro and Teukolsky (CST) \cite{CST} and
by Stergioulas and Friedman \cite{SterFrie} to build numerical models of neutron stars.
%In Sec.~\ref{sec_3}
Next we apply Ryan's method for the BI metric and get a direct relation of the asymptotic term
coefficients with the first multipole moments. At this point we explain why a generic isolated body
in quasi-isotropic coordinates does not have the same asymptotic
metric behavior as the corresponding Schwarzschild metric. Finally %in Sec.~\ref{sec_4}
we discuss the improvement produced in matching the numerical spacetime of a
rotating neutron star with an analytic metric, like the Manko et al. one \cite{manko1},
when the right quadrupole moment of the neutron star model is used in
the analytic metric, instead of the one that was used %up
in the past in similar comparisons. We end up %this section
by giving a short list of other consequences of not using the right multipole moments in
various astrophysical explorations of compact objects.

%%%%%%%%%%%%%%%%%%%%%%%%%%%%%%%%%%%%%%%%%%
\section{\label{sec_2}Asymptotic expansion of a metric}

%%%%%%%%%%%%%%%%%%%%%%%%%%%%%%%%%%%%%%%%%%

In 1976 BI wrote the relativistic equations for the structure and
the gravitational field of a uniformly rotating fluid body. They
assumed that the line element has the following form:
\bea
ds^2 &=& -e^{2\nu} dt^2 + r^2 \sin^2\theta B^2 e^{-2\nu}
(d\phi-\omega dt)^2\nn\\&&+
e^{2(\zeta-\nu)} (dr^2+r^2d\theta^2),
\eea
where $\nu$, $B$, $\omega$, and $\zeta$ are the four metric functions, all
functions of the quasi-isotropic coordinates $r,\theta$ (the other two
coordinates $t,\phi$ do not show up in the metric functions since the
geometry is assumed stationary and axially-symmetric). By writing down
the field equations and the equations of motion for the fluid they
obtained differential equations for the metric functions (see Eqs.~(4-7)
of \cite{ButtIpse}) through which they constructed the asymptotic expansion
of the three metric functions ($\nu,\omega,B$) while the last metric function
$\zeta$ could be easily computed from the rest by a suitable integration.
We copy here these asymptotic expansions since the various coefficients are
intimately related to the multipole moments of the central object as will be
shown later on.

\bea
\nu &\sim&
\left\{-\frac{M}{r}+\frac{\tilde{B}_0M}{3r^3}%+\frac{J^2}{r^4}+\left[-\frac{\tilde{B}_0^2}{5}
               %+\frac{\tilde{B}_2^2}{15}-\frac{12J^2}{5}\right]\frac{M}{r^5}
               +\ldots\right\}+ \left\{\frac{\tilde{\nu}_2}{r^3}+\ldots\right\}P_2
    +\ldots , \label{first}\\
\omega &\sim&
\left
[\frac{2J}{r^3}-\frac{6JM}{r^4}+\left(8-\frac{3\tilde{B}_0}{M^2}\right)\frac{6JM^2}{5r^5}
                +\ldots\right
                ]\frac{d P_1}{d\mu}\nn\\
         &&+ \left[\frac{\tilde{\omega}_2}{r^5}+\ldots\right]\frac{d P_3}{d\mu}+\ldots , \label{second}\eea
\be B \sim
 \sqrt{\frac{\pi}{2}}\left[\left(1+\frac{\tilde{B}_0}{r^2}\right)T_0^{1/2}
               %  \left(\frac{\pi}{2}\right)^{1/2}
                 +\frac{\tilde{B}_2}{r^4}T_2^{1/2}+\ldots\right]. \label{third}
\ee
In the formulae above $P_l$ %and $P_{l ,\mu}(\mu)$
are the Legendre polynomials expressed as functions of $\mu=\cos\theta$, $T^{1/2}_l$ are the so called Gegenbauer
polynomials (similar to the Legendre polynomials, also functions of $\mu$), and $M,J$ are the first two
multipole moments (the mass and the spin) of the spacetime. The rest coefficients are related to the higher
multipole moments.

In 1989 KEH  proposed a different scheme for integrating the field equations using Green's functions.
The line element they assumed was a bit different than the previous one:
%\be
\[ds^2\!=\! -e^{2\nu}dt^2
+ r^2 \sin^2\theta e^{2 \beta}(d\phi-\omega dt)^2
+ e^{2\alpha}(dr^2+r^2d\theta^2).
\]%\ee
The new metric functions  are related to the metric functions of BI
by the following simple relations:
\be
\nu_{\textrm{BI}}=\nu_{\textrm{KEH}}=\nu,~ B_{\textrm{BI}} e^{-\nu}=e^{\beta_{\textrm{KEH}}}~,~ \zeta_{\textrm{BI}}=\nu+\alpha_{\textrm{KEH}}.
\ee
The combinations of $\nu_{\textrm{KEH}}$ and $\beta_{\textrm{KEH}}$
\be
\gamma=\nu_{\textrm{KEH}}+\beta_{\textrm{KEH}}~,~\rho=\nu_{\textrm{KEH}}-\beta_{\textrm{KEH}},
\ee
along with $\omega$ could be expressed as power series in $1/r$, in the same
manner as in Eqs.~(\ref{first},\ref{second},\ref{third}):
\bea
\rho &=& \sum_{n=0}^{\infty}(-2\frac{M_{2n}}{r^{2n+1}}+\textrm{higher order}) P_{2n}(\mu),\\
\omega&=&\sum_{n=1}^{\infty}(-\frac{2}{2n-1}\frac{S_{2n-1}}{r^{2n+1}}+\textrm{higher order})\frac{P_{2n-1}^1(\mu)}{\sin\theta},\\
\gamma&=&\sum_{n=1}^{\infty}(\frac{D_{2n-1}}{r^{2n}}+\textrm{higher order})\frac{\sin(2n-1)\theta}{\sin\theta}.
\eea
In Ryan's 1997 paper \cite{Ryan97} the coefficients $M_{2n}$ and $S_{2n-1}$
were identified as the mass and current-mass moments, respectively, of the
corresponding spacetime; this statement is not exact as will be explained in the next section.
By a simple comparison between the above expansion and the corresponding ones of BI
we see that $M_{2}=-\tilde{\nu}_2$ and $S_3=\frac{3}{2} \tilde{\omega}_2$.

The numerical scheme of KEH was applied in the
numerical codes of \cite{CST,SterFrie} which were then used by various people in order to
construct realistic models of neutron stars; the values $M_{2n}$, $S_{2n-1}$ of these
numerical neutron stars  were then read from the coefficients of the above asymptotic expansions.

%%%%%%%%%%%%%%%%%%%%%%%%%%%%%%%%%%%%%%%%%%
\section{Identifying the multipole moments through Ryan's method}
\label{sec_3}
%%%%%%%%%%%%%%%%%%%%%%%%%%%%%%%%%%%%%%%%%%

In 1995 Ryan wrote coordinate-independent expressions that
relate the energy of a test body that is orbitting in the stationary and axially
symmetric spacetime of an isolated compact body
along a circular equatorial orbit, to the Hansen-Geroch multipole moments
of the spacetime itself \cite{Ryan}.
We wrote similar expressions for the asymptotic expansions
of the metric functions of Eqs.~(\ref{first},\ref{second},\ref{third}) and then related
the series coefficients to the multipole moments of the corresponding
spacetime. Following the procedure of Ryan we first
wrote the orbital frequency of $\Omega$ of the test body,
through the dimensionless parameter $v=(M \Omega)^{1/3}$
as a power series in $x=(M/r)^{1/2}$ and then inverted it
to obtain $x$ as a series in $v$. Following the same procedure, we then calculated the
energy per mass $\tilde{E}$ of a test particle in a circular equatorial orbit, as a function of
$x$ (the expression for $\tilde{E}$ as a function of the metric is given in Eqs.~(10,11) of \cite{Ryan}).
Finally, the energy change per logarithmic interval of the rotational frequency
$\Delta \tilde{E}=-d\tilde{E}/d \log{\Omega}$ was expressed as
a power series in $v$, with coefficient terms written as polynomials of the metric coefficients:
\bea
\Delta\tilde{E} &=&
\frac{\upsilon^2}{3}-
\frac{\upsilon^4}{2}+
\frac{20 j \upsilon^5}{9}-
\frac{(89 +32b +24q)}{24} \upsilon^6+
\frac{28 j\upsilon^7}{3}  \nn\\
&& - \frac{5\left(1439+ 896b- 256j^2+ 672 q\right) \upsilon^8}{432}\nn\\
&& + \frac{\left((421+64b-60q)j-90w_2\right) \upsilon^9}{10}
+O(\upsilon^{10}),
\label{DE}
\eea
where $j=J/M^2,$ $q=\tilde{\nu}_2/M^3,$ $w_2=\tilde{\omega}_2/M^4,$ $b=\tilde{B}_0/M^2$.
By equating the coefficients of the previous power series with the
corresponding ones of Ryan (Eq.~(17) of \cite{Ryan})
we yield directly the right relations between
the coefficients of BI (or of KEH) and the multipole moments
of the spacetime. More specifically from the coefficients of $v^6$ and $v^9$ terms of
the two series, we yield the following values for
the quadrupole mass-moment and the octupole
current mass-moment:
\bea \label{correct_M2}
M_2^{GH} &=& %-\tilde{\nu}_2 - \frac{4}{3} (\frac{1}{4}+b) M^3
 M_2-\frac{4}{3}\left(\frac{1}{4}+b\right)M^3,\;% M^3
%,
%
\\
S_3^{GH}&=& %\frac{3}{2} \tilde{\omega}_2 - \frac{12}{5} (\frac{1}{4}+b) j M^4
S_3-\frac{12}{5}\left(\frac{1}{4}+b\right)jM^4%J M^2
,
\label{correct_S3}
\eea
respectively, where $M_2$ and $S_3$
are the multipole moments as they were mistakenly identified by \cite{Ryan97}
and later used by various authors to read the corresponding moments of numerical models.
By replacing these values in the terms of order $v^7$ and $v^8$ of Ryan's Eq.~(17)
we recover exactly the corresponding term coefficients of Eq.~(\ref{DE}), as expected.
Henceforth we will omit the superscript $^{GH}$ in $M_2,S_3$ when we refer to the right
Hansen-Geroch multipole moments.

The last terms in Eqs.~(\ref{correct_M2},\ref{correct_S3}), which were missing up to now
in the literature, are the ones that are causing the discrepancy between the
estimated incorrect moments and the true moments. Both these extra terms
would be zero for $b=-1/4$. This was pointed out by Laaarkkers and
Poisson \cite{LaarPois}. These correcting terms were considered harmless though, since
the Schwarzschild metric corresponds to $B=1-M^2/4 r^2,$ that is to $b=-1/4$. This,
was argued by Laarakkers and Poisson, should correspond to the lowest order term of the metric function $B$,
(the term in front of $T^{1/2}_0$ in Eq.~(\ref{third})) of any axisymmetric isolated body.
This is not true though, if $r$ is the isotropic coordinate radius.
The lowest order asymptotic term of any metric describing a stationary isolated object
is simply the 1 in the first order term of $B$; the higher orders
are generally expected to deviate from their Schwarzschild corresponding terms.

In the recent book of Friedman and Stergioulas \cite{Book} this discrepancy was noted and was corrected by a
transformation of the $r$ coordinate, leading to exactly the same correction for $M_2$ as in our Eq.~(\ref{correct_M2}).
The analysis in our paper though is general and coordinate-independent.
It could be used to treat any set of coordinates and find the relation between the corresponding coefficients and the true
multipole moments. Moreover, in our paper we give the right formula for $S_3$ as well.

As a final remark, we note that the Kerr metric, expressed in
isotropic coordinates \cite{Lanz}, yields a $B$ function, $B=1-(M^2-a^2)/4 r^2,$ hence $b^{Kerr}=-(1/4)(1-j^2)$. This
result is a clear manifestation of the erroneous assumption that all stationary axisymmetric metrics correspond
to $b=-1/4$.

%%%%%%%%%%%%%%%%%%%%%
\section{Consequences of evaluating
the right moments}
\label{sec_4}
%%%%%%%%%%%%%%%%%%%%%

We now present a list of the various effects caused by
computing correctly the multipole moments of the numerical models of
neutron stars, and give a short account, whenever this was possible, of the subsequent
quantitative alterations in recent scientific conclusions related to studies of the exterior field of
neutron stars.

(i) We start with the attempt of constructing analytic vacuum
solutions of Einstein's equations that could then be used to fit the
various numerical models of rotating neutron stars. Berti and Stergioulas (BS) \cite{BertSter} tried to
match a three-parameter analytic solution \cite{manko1} to a wide diversity of uniformly
rotating neutron-star models. Each analytic solution was
constructed so that its first three multipole moments were equal to
the corresponding moments (mass, spin and quadrupole) of the
particular neutron star, where these moments were read directly from the
asymptotic expansion of the corresponding numerical metric. Their
conclusion was that this type of analytic solution was quite good
to describe the external metric of all kinds of fast
rotating neutron stars. Since the specific metric cannot assume low values of quadrupole moment, the
metric is not adequate to describe the slowly rotating neutron stars. Whenever an analytic
solution could be constructed, the matching
was such that the two metrics (analytic and numerical)
did not differ by more than $6 \%$ at the surface of the star.

\begin{figure}
\includegraphics[width=0.4\textwidth]{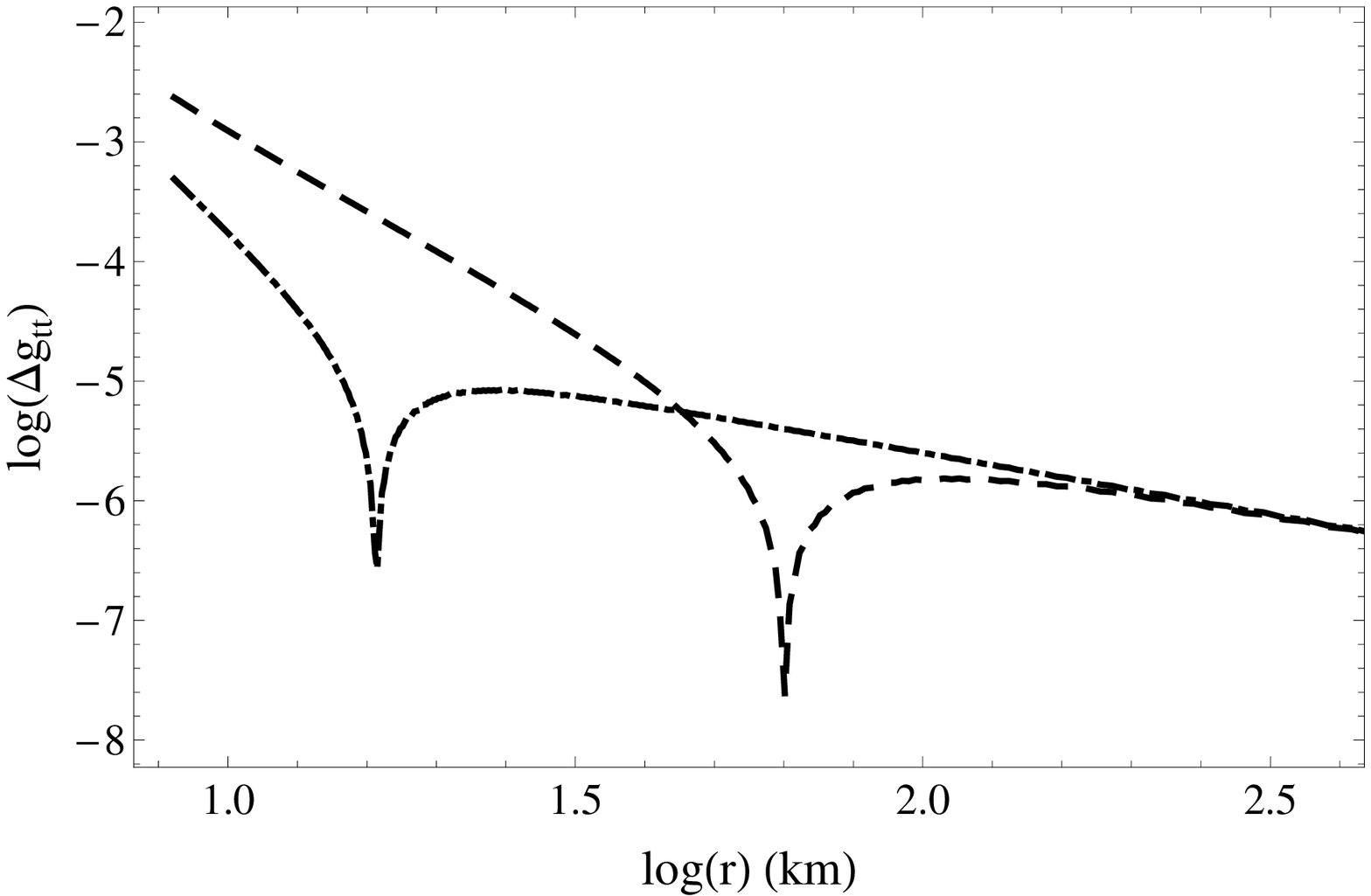}
\includegraphics[width=0.4\textwidth]{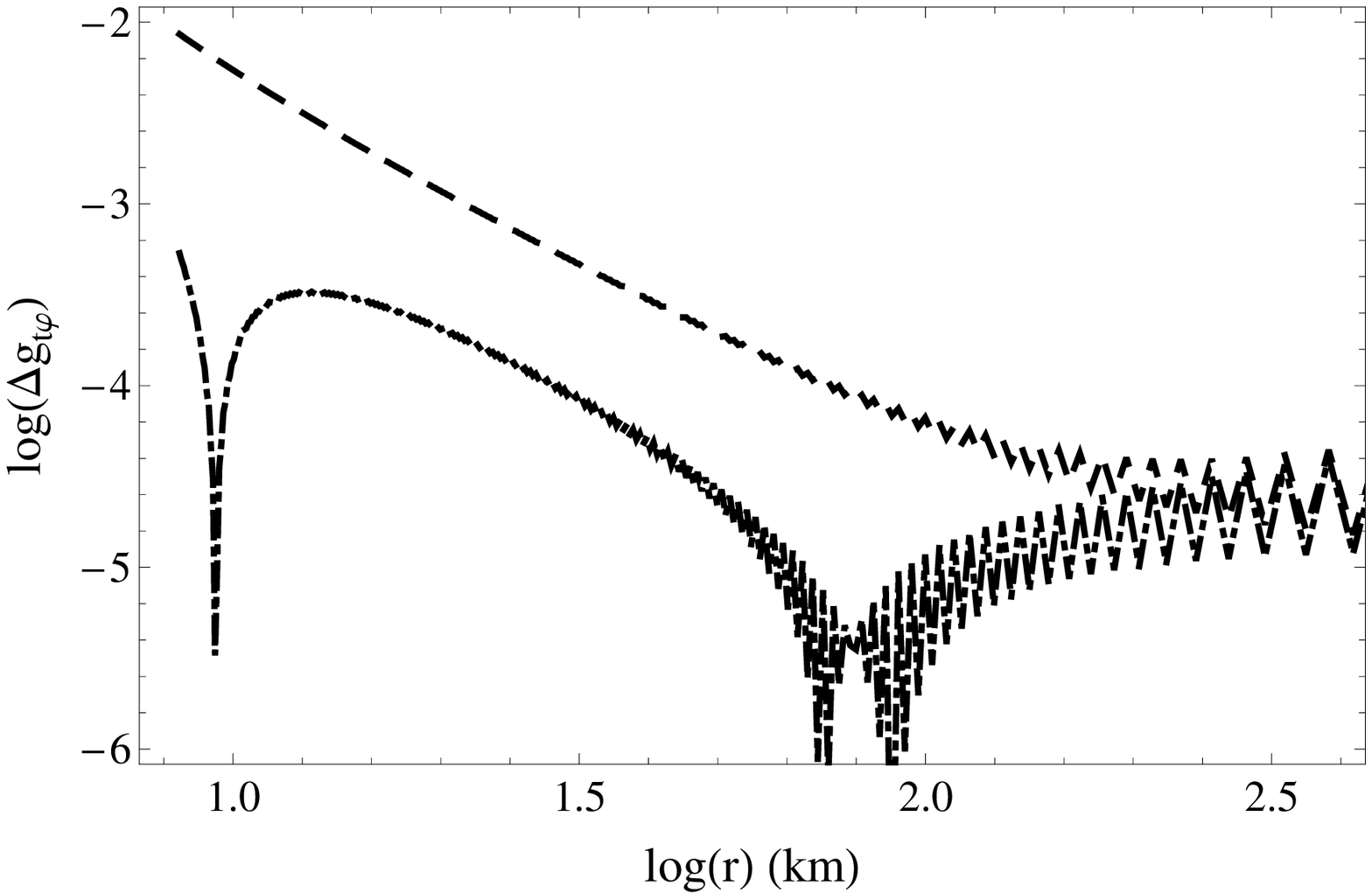}
\protect
\caption{A typical log-log plot of the relative difference between the
numerical and the analytic metric ($(g_{ij}^n-g_{ij}^a)/g_{ij}^n$) for a specific numerical model
(model $\# 16$ of EOS FPS of \cite{complement}), % at various distances from the star,
before (dashed curve) and after the correction of $M_2$ (dashed-dotted curve).
%The deep wells are simply due to crossings between the analytic and the numerical metric, while the
%oscillations of the tail of the bottom plot are due to numerical errors of the numerical metric.
The top plot is for $g_{tt}$ and the bottom one for $g_{t \phi}$. % In order to relate the graphical
%mismatch with the measure of improvement used in the text,
We note that the corresponding overall improvement for this particular neutron star model was 6.6 in $g_{tt}$
and 15.1 in $g_{t \phi}$. This was a model with a medium improvement in $g_{tt}$, compared to
the whole set of models which were examined.}
\label{fig}
\end{figure}

We attempted the same comparison as in \cite{BertSter}, assuming the same analytic spacetime and using all numerical
models with EOS's AU, FPS and L, while the quadrupole moment that was inserted in
the analytic solution was the one corrected according to Eq.~(\ref{correct_M2}).
In order to correct it, we used the asymptotic expansion of the metric function $B$ of the numerical metric.
Although the quadrupole moment was not affected by more than $\sim 20 \%$, and for some numerical models by a much
lower fraction, the improvement it evoked in matching the numerical metric was almost an order of magnitude.
We computed the overall mismatch between the numerical and the analytic metric exterior to the star,
which was defined as $\sigma_{ij}=(\int_{R_S}^{\infty}(g_{ij}^n-g_{ij}^a)^2dr)^{1/2}$,
where $n,a$ indicate the numerical and the analytic metric components respectively and $R_S$ is the surface radius.
The improvement we gained in the overall mismatch was a factor of $\sim 2$ to $\sim 8$ for the
$g_{tt}$ and $\sim 2$ to $\sim 15$ (in most cases) for the $g_{t\phi}$. In a few cases the improvement in the mismatch of
$g_{t \phi}$ (and in one model of $g_{tt}$) was either marginal or
worse after the correction of $M_2$. This happened because, while we correct the $M_2$ of the Manko et al. metric,
its octupole $S_3$ is altered in such a way that the analytic metric finally raises its overall mismatch.
However, even these not improved cases have an extremely good overall mismatch
(less than 0.004 after the worsening). For a full presentation of the numerical results of the
mismatch and the fractional deviation of $M_2$ and $S_3$ due to correction, see \cite{complement}.
Figure \ref{fig} shows an example of the mismatch between analytic and numerical metric functions before and
after the correction.

%%%%%%%%%%%%%%%%%%%%%%%%%%%%%%%%%%%%%%%%%%%%%%%%%%%%%%%%%%%%%%%%
%%%%%%%%%%%%%%%%%%%%%%%%%%%%%%%%%%%%%%%%%%%%%%%%%%%%%%%%%%%%%%%%

Whenever an analytic metric is used to mimic the gravitational field of a realistic neutron star \cite{inprep},
and then this metric is used to reproduce various observables related to neutron stars \cite{Pappas}, it is important
to use the right moments in order to build a faithful analytic metric; otherwise any physical conclusion inferred
by the analytic metric would be off \cite{Pach11}. If quantitative conclusions are drawn while using wrong
moments in the analytic metric, they should be taken into account with some reservation.

(ii) Another important effect of altering the values of the multipole moments
of numerical neutron-star models is in relating the higher moments of the compact object with its spin $j$.
This was attempted by Laarakkers and Poisson \cite{LaarPois} for the quadrupole moment, but due to
the omission of the $b$-related term the parameters estimated when
relating $M_2$ to $j^2$ were somewhat off. We repeated their analysis with
the corrected values of $M_2$ and we got a bit different fitting parameters.
In \cite{complement} one could find our parameters; however they cannot not be
directly compared to the ones of \cite{LaarPois} since the sequences of model we used are
not exactly equivalent to the ones of \cite{LaarPois}. Having the right relations between various moments
could help us interpret future accurate observations with connection to the internal structure of neutron stars.
At this point we should note that the correcting factor $b$ deviates from $-1/4$ following a quadratic relation
with $j$ as well, thus preserving the quadratic fit of $M_2$ with $j$ that was found in \cite{LaarPois}.
Furthermore we found a similar empirical relation for $S_3$ with $j$, namely $S_3=a_3 j^3$,
with $a_3$ a constant parameter depending on the EOS and the mass of the neutron star (see \cite{complement}).

(iii) The ISCO radius, which is significant for a number of astrophysical observations, is intimately related to
the exact relativistic moments of the central object, since it lies close to it.
In \cite{Bert} the ISCO of  various numerical models of neutron stars were compared to the ISCO computed
for a Hartle-Thorne (HT) analytic metric \cite{HT}, that was suitably constructed to match the behavior of
the numerical models. What they found was that although the relative difference in quadrupole between HT and
numerical models was oddly high even at slow rotations (where one would expect almost perfect match), the corresponding
ISCOs were extremely close (at low spins). This intriguing disagreement disappears for the right quadrupoles
of the numerical models. On the other hand the ISCOs were originally in good agreement since they are not affected by
such misidentification of $M_2$ and were accurately computed for both metrics. However, that would not be
the case if the approximate formula of SS % Shibata and Sasaki
\cite{ShibSasa} was used to compute the ISCO, since it is a function of the various moments.
We repeated the comparison of ISCOs \cite{BertSter} between the numerical one and that obtained from SS and we
found a partial improvement but not an impressive one. The reason is that
at low quadrupoles (corresponding to low spins) SS's formula
is extremely accurate and quite insensitive to small corrections of $M_2$,
while at high quadrupoles the correction of ISCO through $M_2$ is significant
but then the formula deviates a lot from the true ISCO. For example
the relative difference of ISCO drops from $0.92 \%$ to $0.61 \%$ (due to
correction of $M_2$) for low $M_2$, while it drops from $17.3 \%$ to $15.6 \%$ for a large
value of $M_2$ (for a sequence of models of FPS).

(vi) Finally, in \cite{Psal} the apparent surface area of a rotating
neutron star, due to its quadrupole deformation, is computed as a
function of this deformation. The deformation parameter though is
read from the quadratic fit of \cite{LaarPois} which is somewhat
distorted due to wrong identification of moments. Hence, the numbers
are once again slightly deviated from their true values.

%%%%%%%%%%%%%%%%%%%%%%%%%%%%%%%%%%%%%%%%%%%%%%%%%%%%%%%%%%%%%%%%%%%%%%%%%%%%%%
%%%%%%%%%%%%%%%%%%%%%%%%%%%%%%%%%%%%%%%%%%%%%%%%%%%%%%%%%%%%%%%%%%%%%%%%%%%%%%
%%%%%%%%%%%%%%%%%%%%%%%%%%%%%%%%%%%%%%%%%%%%%%%%%%%%%%%%%%%%%%%%%%%%%%%5555555

\section*{Acknowledgements}
We would like to thank N.~Stergioulas and K.~Kokkotas for helpful
and enlightening discussions on the subject. GP would like to thank
H. Markakis for the discussions that led us search more deeply the apparent
disagreement between numerical models and analytic metrics. The work
was supported by the research funding program of I.K.Y. (IKYDA
2010).

%\bibliography{MM_pappas}

%\section*{References}

%\pagebreak
\clearpage

%------------------------------------------------------------------------
\section{Supplemental material}
%------------------------------------------------------------------------

%------------------------------------------------------------------------
\subsection{Preliminary remarks}
\label{sec1}
%------------------------------------------------------------------------

In this supplement, we give a detailed account of all data related
to the neutron star models that we used in our analysis and all
the results from the comparisons between using the previously
assumed moments and the corrected ones.

For our analysis we have constructed several numerical neutron
star models, using Stergioulas' ``rns'' code \cite{SterFrie}.
Specifically, we have constructed the same models as the ones
presented in \cite{BertSter} for the equations of state AU, FPS
and L. These models correspond to sequences of constant baryonic
mass with varying rotation. In particular we have: (i) one
sequence which corresponds to a non-rotating model of mass
$1.4M_{\odot}$, (ii) one sequence which corresponds to the
non-rotating model of maximum mass for the particular equation of
state, and (iii) a sequence that doesn't have a non-rotating
limit. Every sequence consists of 10 models, so we have studied 30
models for every equation of state.

%%%%%%%%%%%%%%%%%%%%%%%%%%%%%%%%%%%%%%%%%%%%%%%
\subsection{Fitting parameters for higher multipoles vs spin}
\label{sec:bestfit}
%%%%%%%%%%%%%%%%%%%%%%%%%%%%%%%%%%%%%%%%%%%%%%%

\begin{figure*}

  \includegraphics[width=0.32\textwidth]{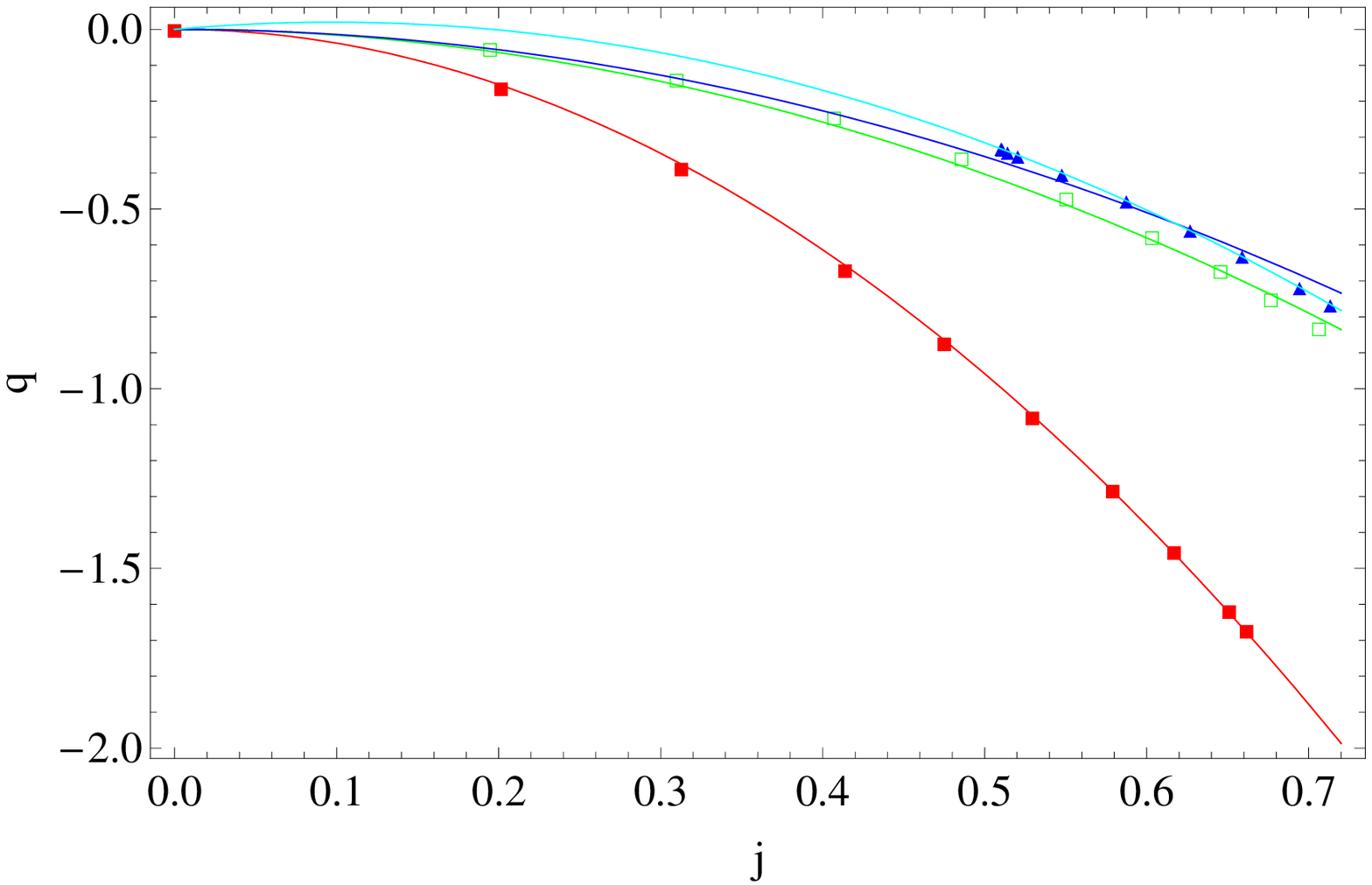}
  \includegraphics[width=0.32\textwidth]{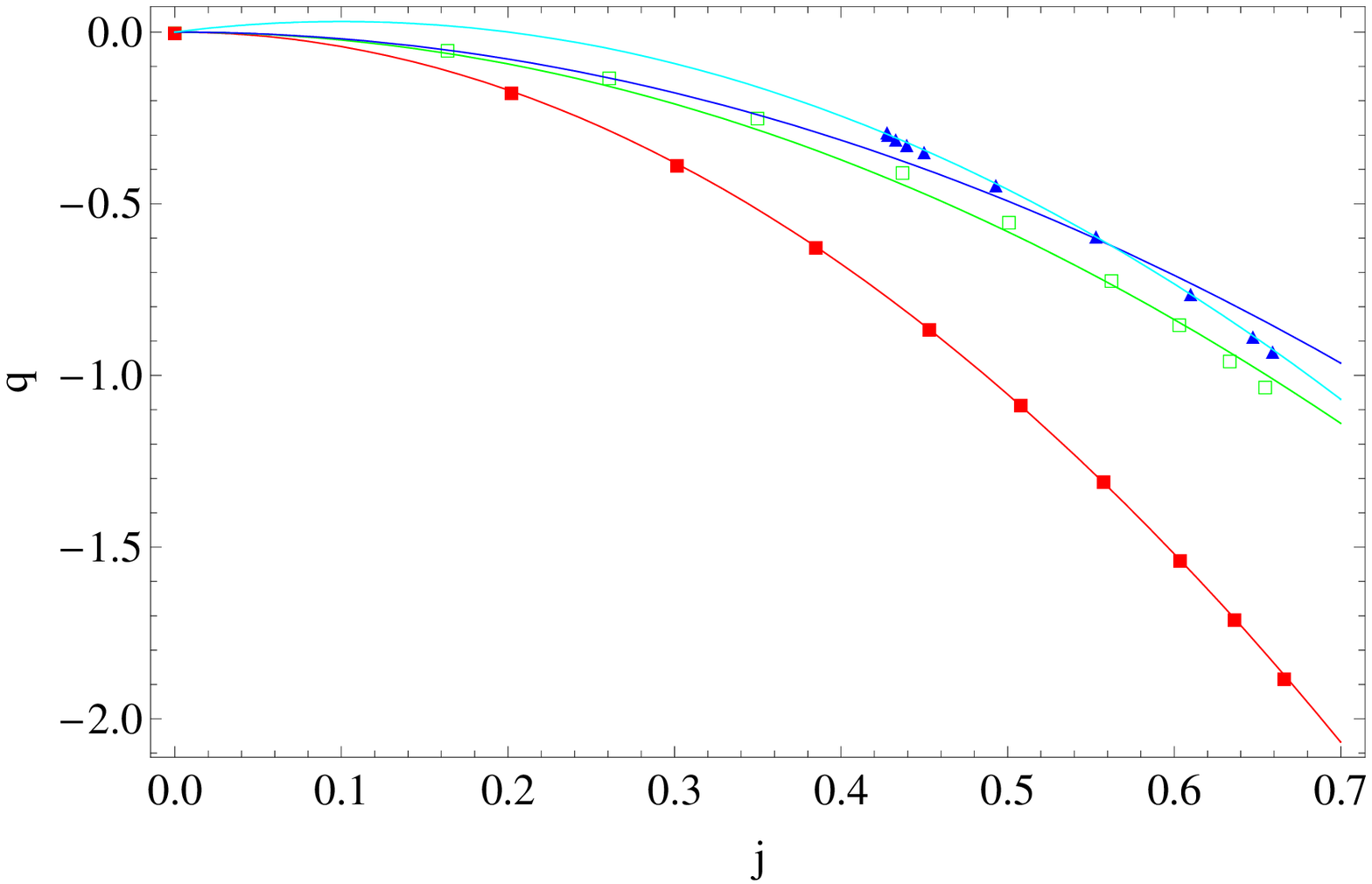}
  \includegraphics[width=0.32\textwidth]{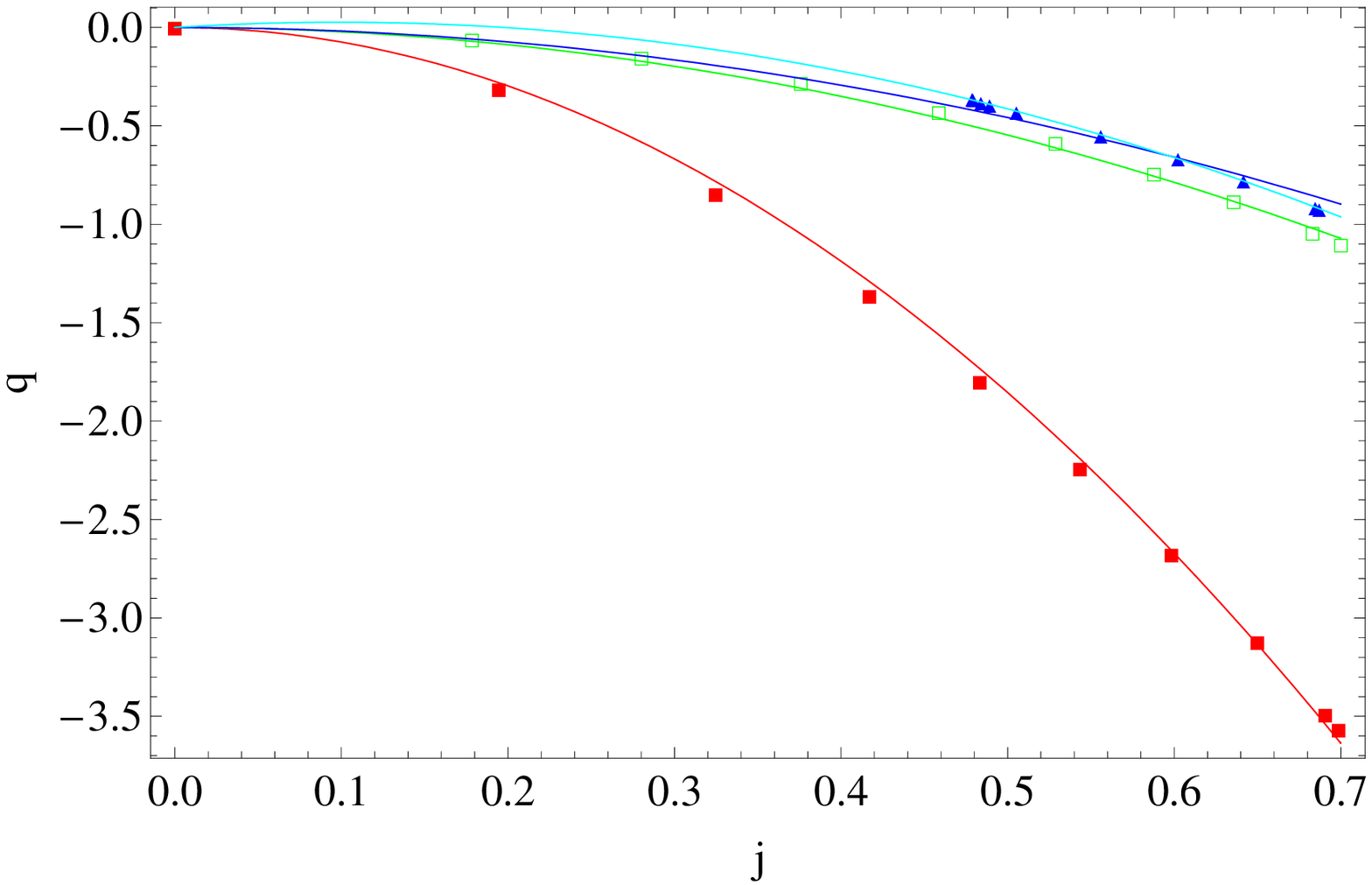}
  \includegraphics[width=0.32\textwidth]{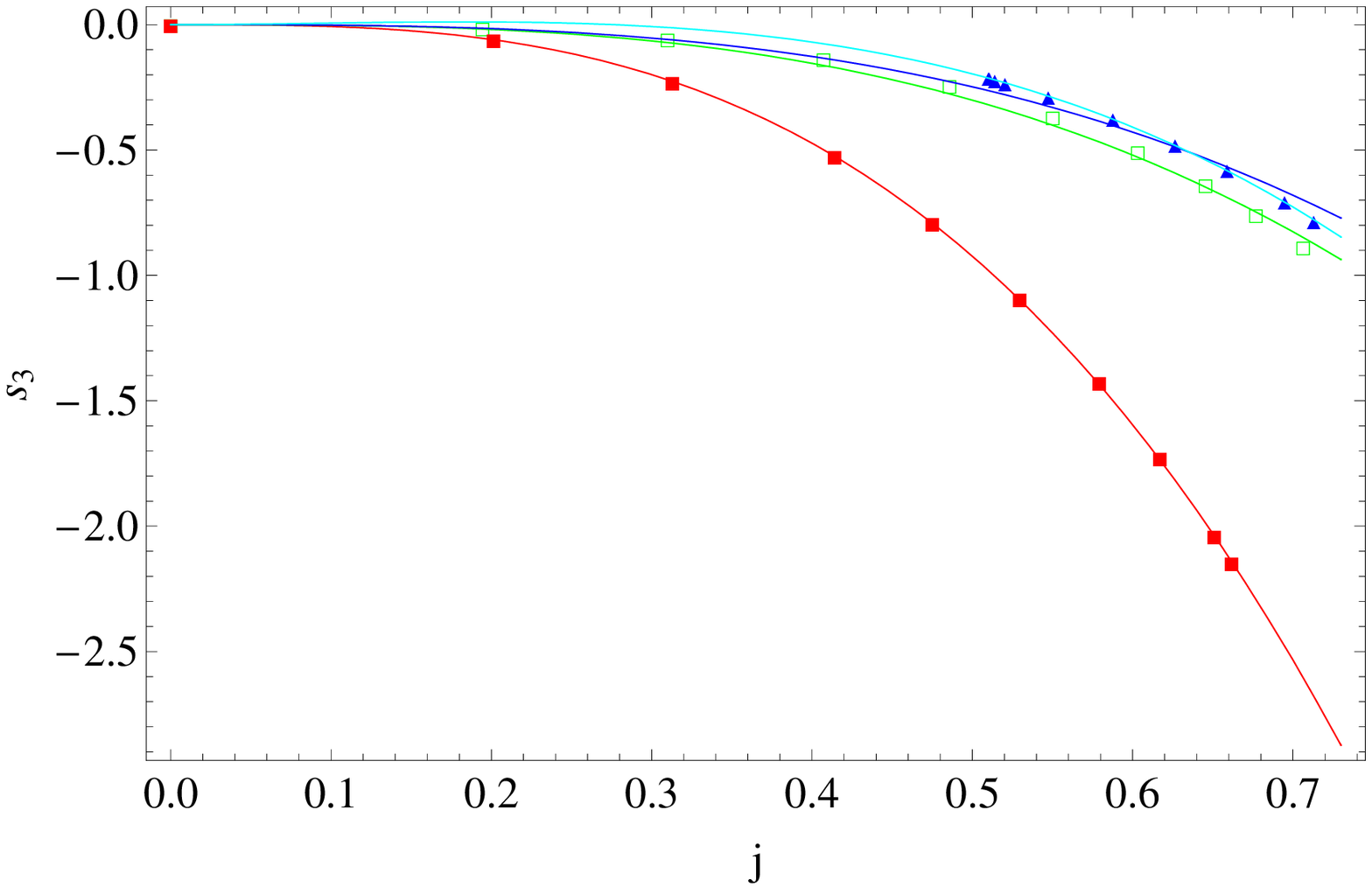}
  \includegraphics[width=0.32\textwidth]{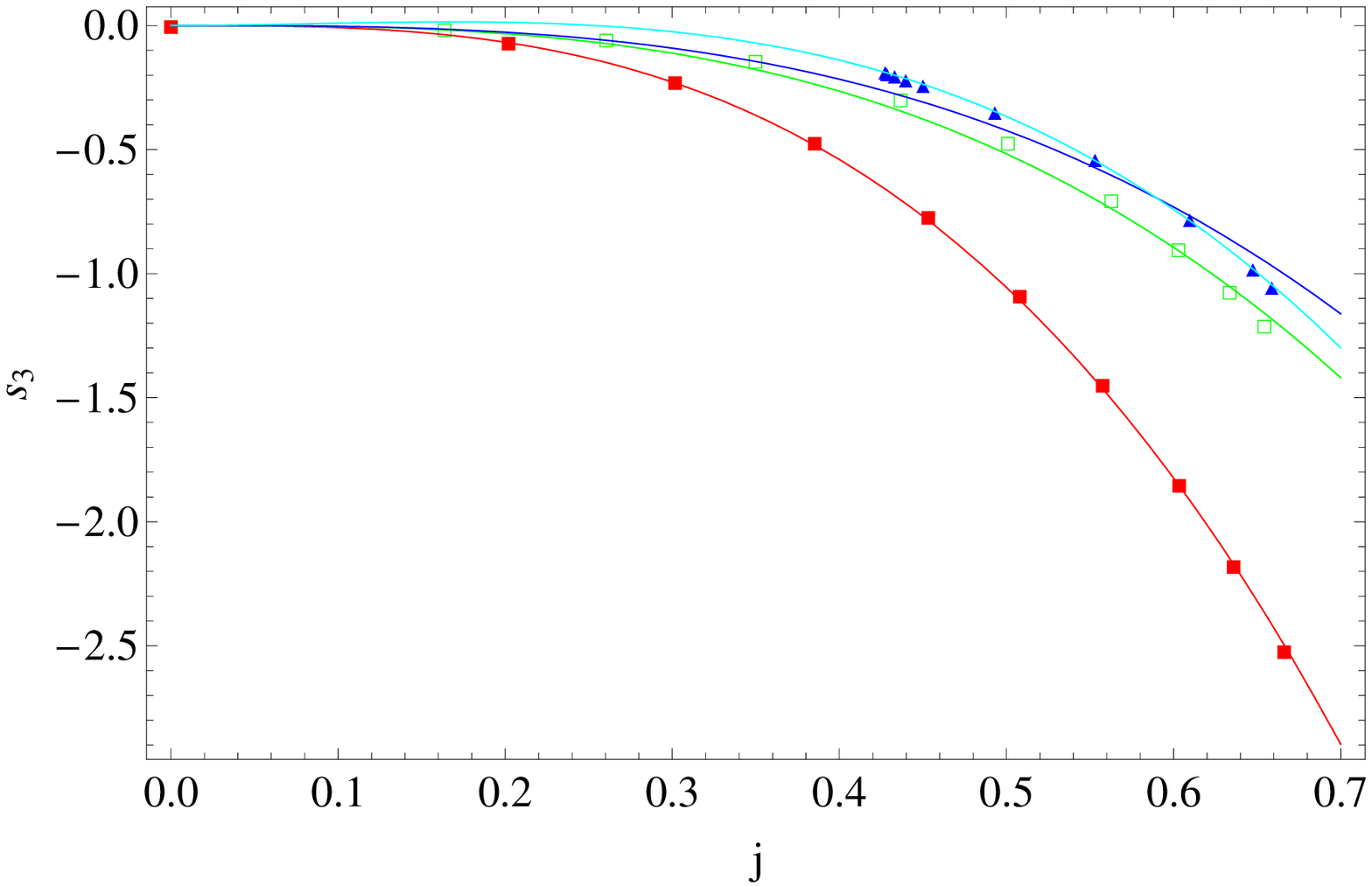}
  \includegraphics[width=0.32\textwidth]{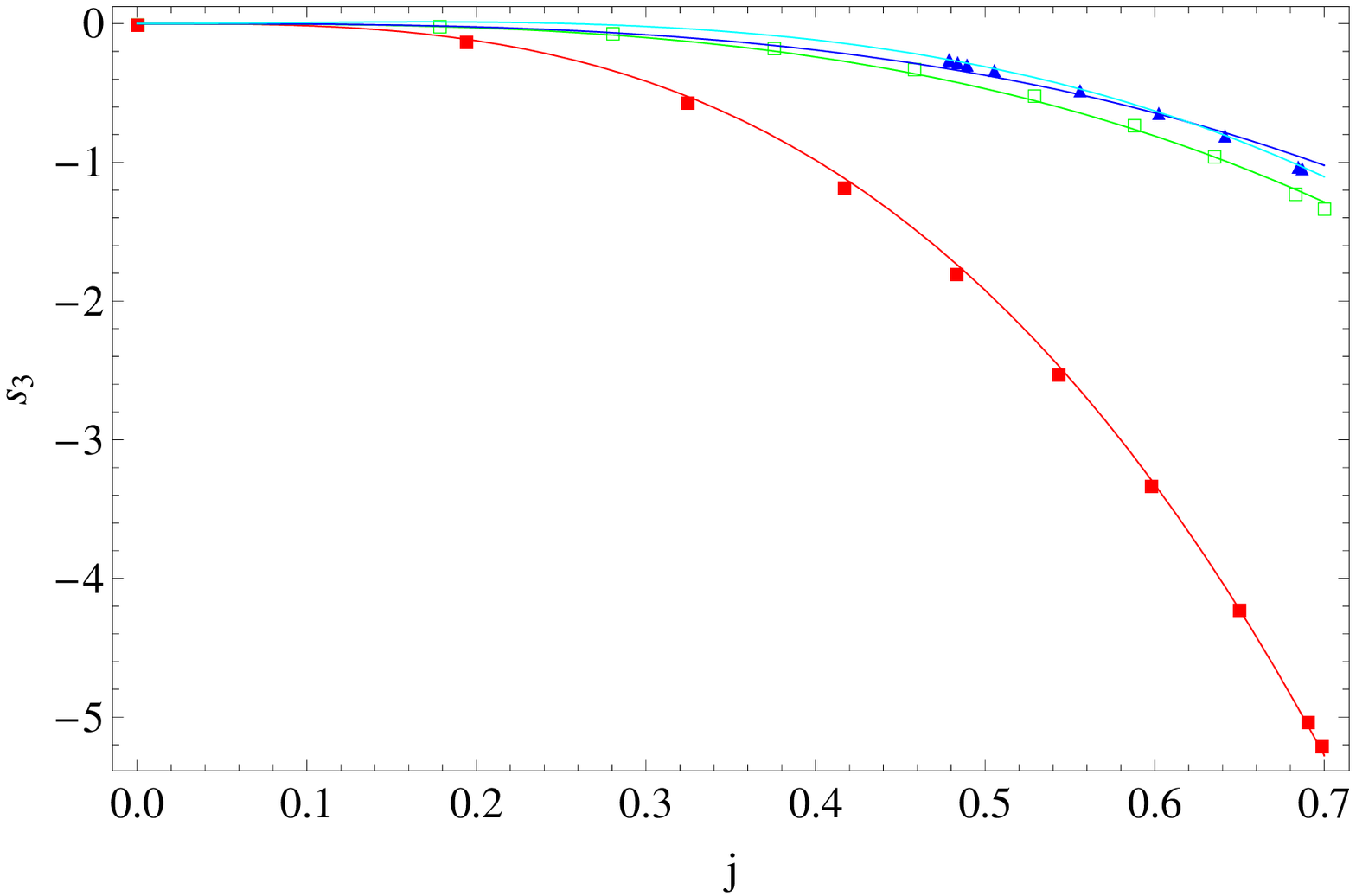}
  \includegraphics[width=0.32\textwidth]{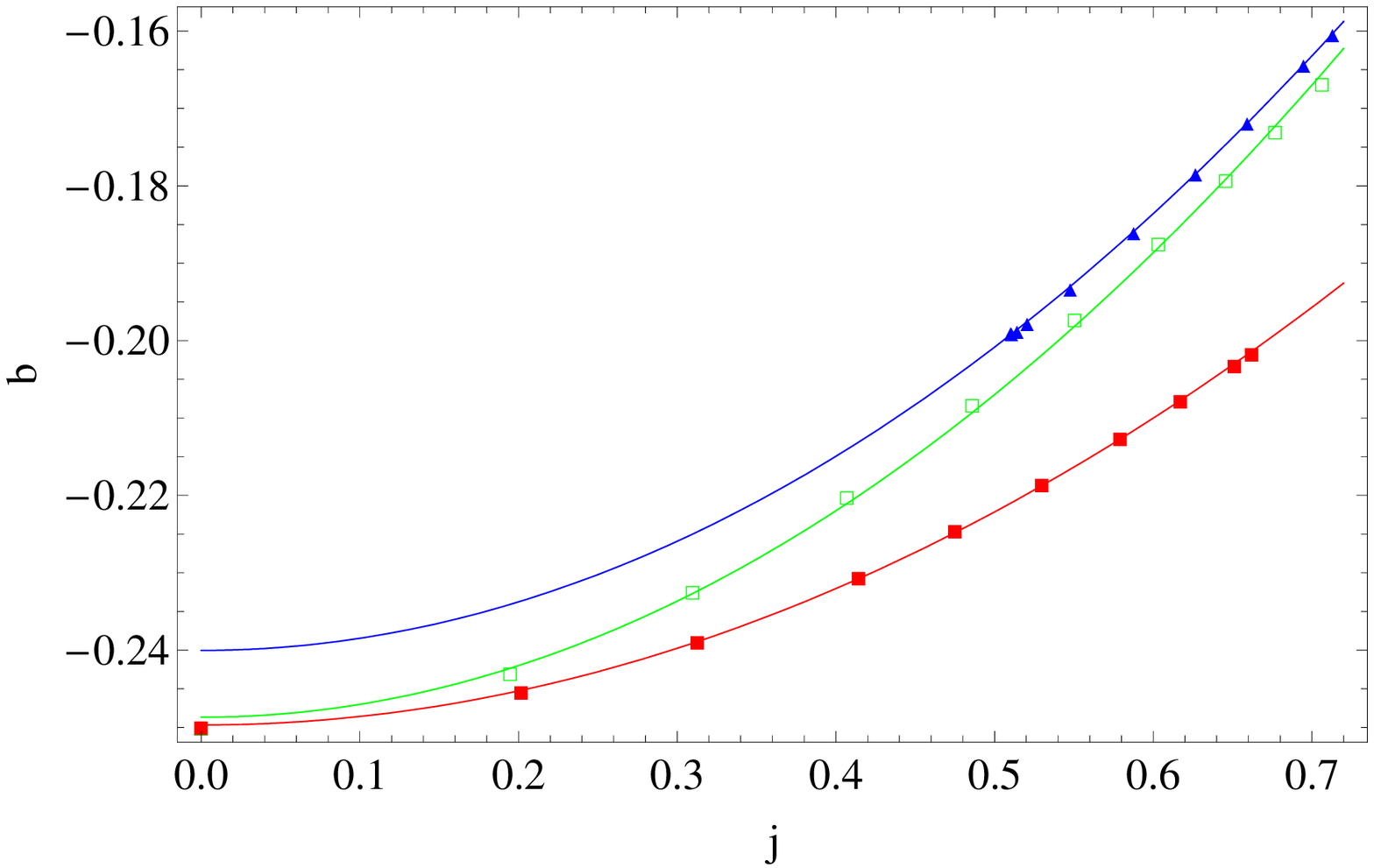}
  \includegraphics[width=0.32\textwidth]{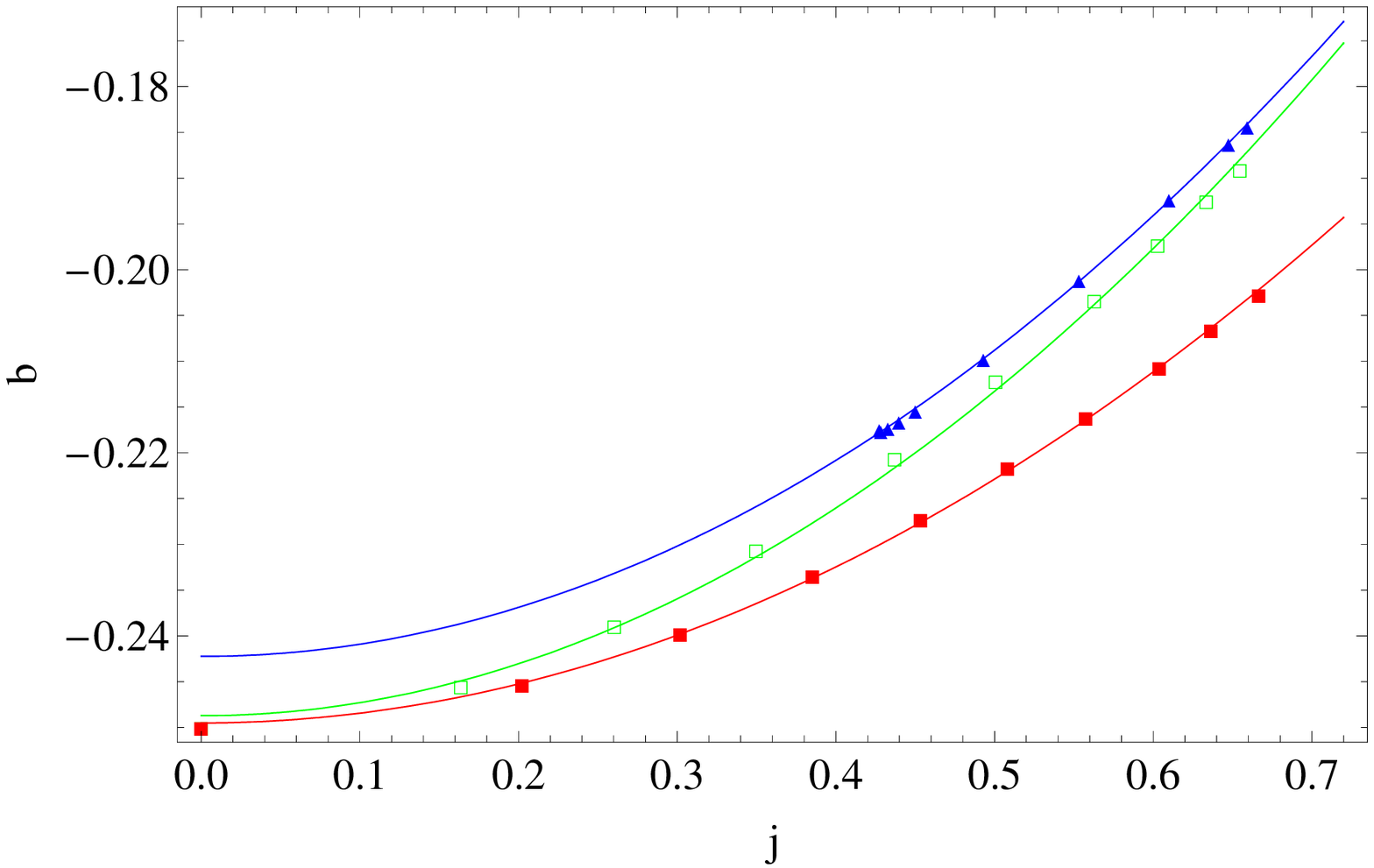}
  \includegraphics[width=0.32\textwidth]{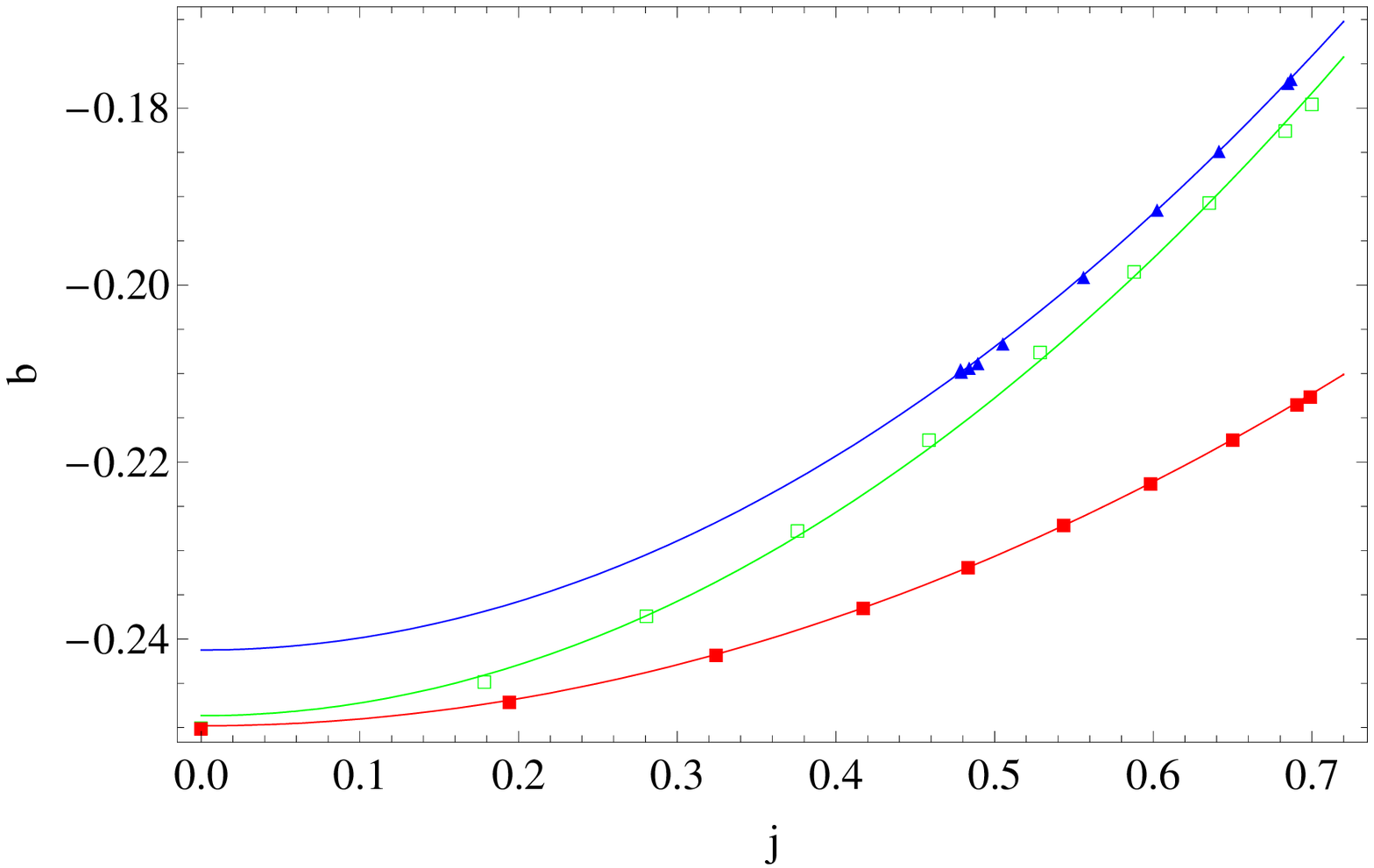}
  \caption{Plot of the reduced moments and the corresponding fits for the three EOSs (AU on the left, FPS in the
  middle and L on the right). In the case of $q$ and $s_3$ (two upper rows), red is the fit for sequence (i),
  green is the fit for sequence (ii), blue is the fit with one
  parameter for sequence (iii), and cyan is the fit with two parameters for the same sequence. The bottom row
  shows the plots and corresponding fits for the parameter $b$ with the same color coding.}
  \protect\label{figsfit1}

\end{figure*}

We have tried to fit the reduced quadrupole mass-moment
$q=M_2/M^3$ and the reduced octupole current-mass moment
$s_3=S_3/M^4$ with simple polynomials of the spin parameter
$j=J/M^2$, for every sequence of the numerical neutron-star models
that were briefly presented in the previous section. The moments
$M_2$ and $S_3$ are obtained from the ``rns'' code and corrected
according to Eq.~(11) of our letter. The dimensionless parameter
$q$ was fitted as a pure quadratic polynomial $a_2 j^2$ of the
spin parameter $j$ for all three sequences, and additionally in
the case of sequence (iii) as a polynomial of the form $a_1 j +
a_2 j^2$ to improve the fit. Accordingly the octupole parameter
$s_3$ was fitted by a pure cubic fit $a_3 j^3$ for all sequences,
and in the case of sequence (iii) as a polynomial of the form $a_2
j^2 + a_3 j^3$. In Fig.~\ref{figsfit1} we present the data for $q$
and $s_3$ as functions of the spin parameter, as well as the
corresponding best-fit polynomials. In Table \ref{tabfit} the
corresponding fitting parameters are shown for each sequence of
models and for each EOS. Furthermore we investigated the behavior
of $b=\tilde{B}_0/M^2$ with $j$, since the $b$ value affects the
correction in $M_2$ and $S_3$ in a linear fashion. We found that
$b$ follows a quadratic relation with $j$ as well; the correction
preserves the quadratic relation of $q$ found in \cite{LaarPois}.
The plots and the corresponding best fits for $b$ are also shown
in Fig.~\ref{figsfit1}, and the fitting parameters are written in
Table \ref{tabfit}.

\begin{table}

 \centering
%\begin{minipage}{1. \textwidth}
  \caption{The moments expressed as polynomials of the spin
  parameter $j$ for the various equations of state: $M_n/M^{n+1}=a_0 +a_1 j +a_2 j^2+a_3 j^3$.
  The top three subtables correspond to the three sequences of
  models. The third sequence of models (the one without non-rotating limit) is also
  fitted with a combination of powers in $j$ (bottom subtable).}
%  \begin{scriptsize}
  \centering
  \begin{ruledtabular}
\begin{tabular}{c c c c c c c c c c}
$a_i$  &        & AU       &       &        & FPS      &       &
& L &
\\
    &   $q$  &  $s_3$  &  $b$  &   $q$  &  $s_3$   &  $b$  & $q$  &  $s_3$ & $b$
        \\\hline
 $a_0$  &    -   &    -    &-0.25  &    -   &    -     &-0.25  &    -   &    -   & -0.25 \\
 $a_1$  &    -   &    -    &   -   &    -   &    -     &   -   &    -   &    -   & -  \\
 $a_2$  & -3.83  &    -    & 0.11  & -4.22  &    -     & 0.11  & -7.42  &    -   & 0.08  \\
 $a_3$  &    -   &  -7.38  &   -   &    -   &  -8.45   &   -   &    -   & -15.38 & - \\\hline
        &   $q$  &  $s_3$  &  $b$  &   $q$  &  $s_3$   &  $b$  &   $q$  &  $s_3$ & $b$ \\\hline
 $a_0$  &    -   &    -    &-0.25  &    -   &    -     &-0.25  &    -   &    -   & -0.25 \\
 $a_1$  &    -   &    -    &   -   &    -   &    -     &   -   &    -   &    -   & -  \\
 $a_2$  & -1.61  &    -    & 0.17  & -2.33  &    -     & 0.14  &  -2.19 &    -   & 0.14  \\
 $a_3$  &    -   &  -2.41  &   -   &    -   &  -4.14   &   -   &    -   & -3.75  & - \\\hline
        &   $q$  &  $s_3$  &  $b$  &   $q$  &  $s_3$   &  $b$  &   $q$  &  $s_3$ & $b$ \\\hline
 $a_0$  &    -   &    -    & -0.24 &    -   &    -     &-0.24  &    -   &    -   & -0.24 \\
 $a_1$  &    -   &    -    &   -   &    -   &    -     &   -   &    -   &    -   & -  \\
 $a_2$  & -1.41  &    -    & 0.16  & -1.97  &    -     & 0.13  &  -1.83 &    -   & 0.14  \\
 $a_3$  &    -   &  -1.98  &   -   &    -   &  -3.39   &   -   &    -   & -2.98  & - \\\hline
        &   $q$  &  $s_3$  &  $b$  &   $q$  &  $s_3$   &  $b$  &   $q$  &  $s_3$ & $b$ \\\hline
 $a_0$  &    -   &    -    & -0.24 &    -   &    -     &-0.24  &    -   &    -   & -0.24 \\
 $a_1$  & 0.40   &    -    &   -   &  0.61  &    -     &   -   &  0.54  &    -   & -  \\
 $a_2$  & -2.07  &  0.96   & 0.16  & -3.06  &   1.5    & 0.13  & -2.74  &  1.29  & 0.14  \\
 $a_3$  &    -   &  -3.50  &   -   &    -   &  -5.93   &   -   &    -   & -5.06  & - \\
\end{tabular}
\end{ruledtabular}
\protect\label{tabfit}
%\end{scriptsize}
%\end{minipage}
\end{table}

%%%%%%%%%%%%%%%%%%%%%%%%%%%%%%%%%%%%%%%%%%%%%%%
\subsection{Matching a neutron-star spacetime with a Manko et al. metric }
%%%%%%%%%%%%%%%%%%%%%%%%%%%%%%%%%%%%%%%%%%%%%%%

We have compared the metric functions $g_{tt}$ and $g_{t \phi}$ on
the equatorial plane of various numerical models of neutron stars
(the same sequences and EOSs presented in Sec.~\ref{sec1}) with
the corresponding Manko et al. \cite{manko1} metric functions. For
every star the analytic metric was constructed with the same mass
$M$, and spin $J$, as the numerical model, while the third
parameter of the analytic solution was adjusted so as its
quadrupole moment equals the quadrupole moment of the numerical
spacetime. That was attempted both when the correction in $M_2$
was omitted (as in \cite{BertSter}), and when the correction was
taken into account. For every numerical model we recorded the
first four non-vanishing multipole moments, as well as the
fractional difference in $M_2$ and $S_3$ due to correction. We
also give the parameter $b$ as evaluated from the numerical
metric. The numbers are shown in Tables \ref{num1}, \ref{num2},
\ref{num3} (the spin parameter $j$ is shown instead of the angular
momentum $J$).

In Tables \ref{anal1}, \ref{anal2}, \ref{anal3} we present the
results of the comparison between the numerical metric functions
and the corresponding analytic ones. More specifically we give:
(a) the overall mismatch $\sigma_{tt}$ and $\sigma_{t\phi}$ with
the corrected value for $M_2$, (b) the improvements in the overall
mismatch due to correction,
\[\frac{\sigma_{tt}|_{old}}{\sigma_{tt}|_{new}},\;\;\frac{\sigma_{t\phi}|_{old}}{\sigma_{t\phi}|_{new}}, \]
and (c) the fractional difference in $S_3$ of the Manko et al.
metric (which, for this particular analytic spacetime, depends on
the first three multipole moments) due to the correction in $M_2$.
We should note that in the case of the Manko et al. metric, the
quadrupole can not take arbitrarily low values; thus not every
numerical model has a corresponding analytic model. We indicate
the specific models in Tables \ref{anal1}, \ref{anal2},
\ref{anal3} by their numbering in Tables \ref{num1}, \ref{num2},
\ref{num3}, respectively.

\begin{figure*}

  \includegraphics[width=0.32\textwidth]{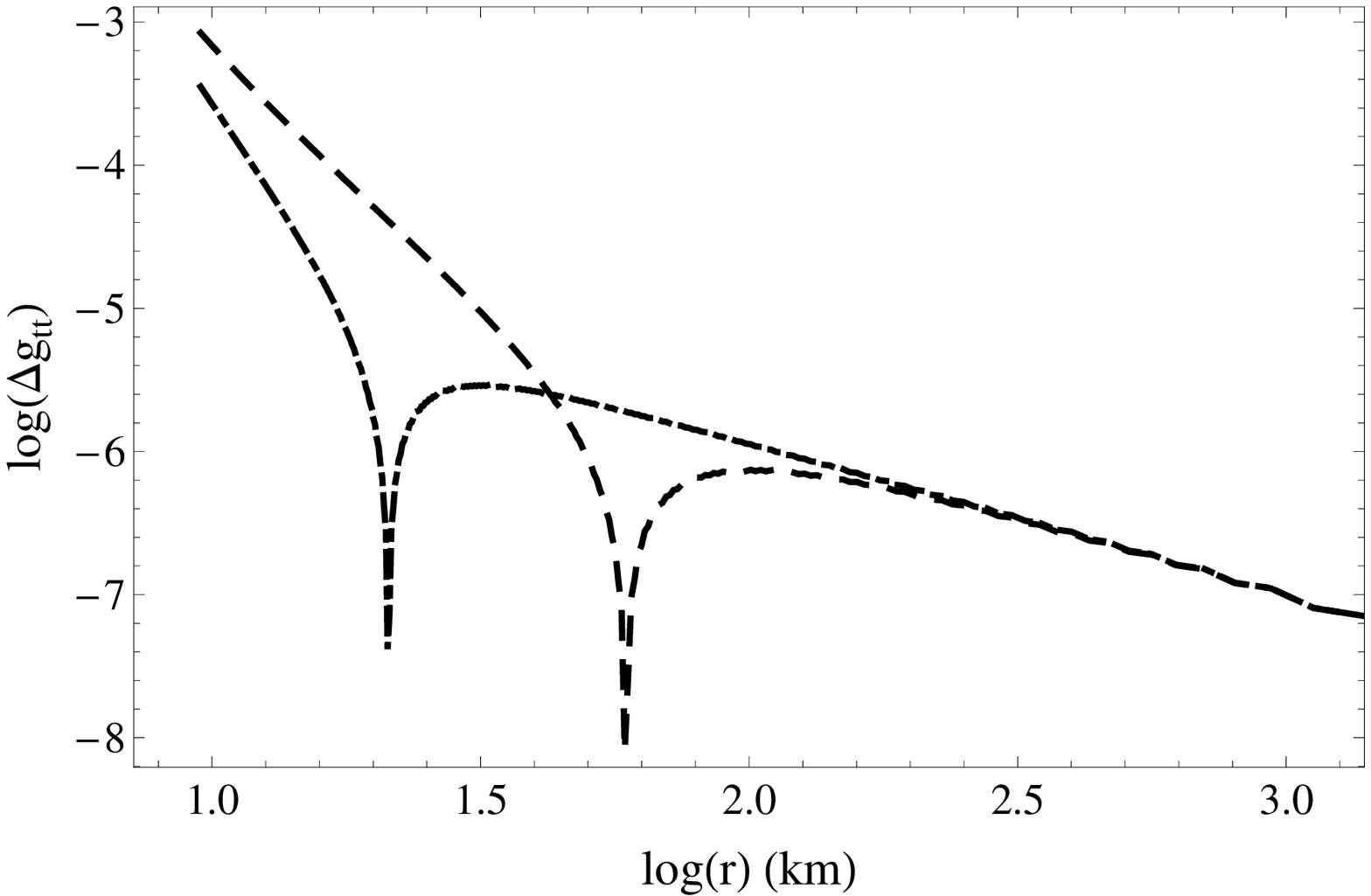}
  \includegraphics[width=0.32\textwidth]{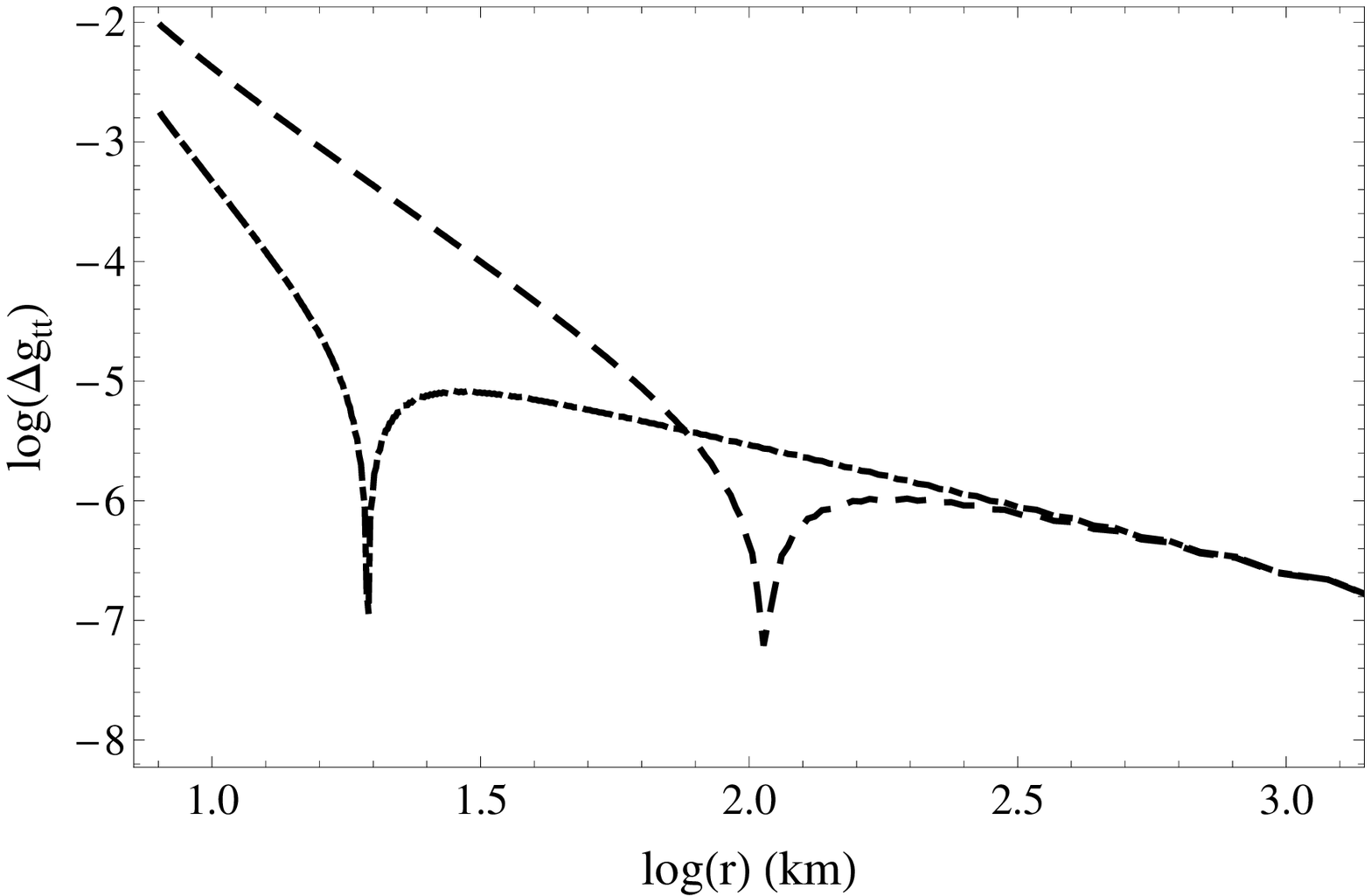}
  \includegraphics[width=0.32\textwidth]{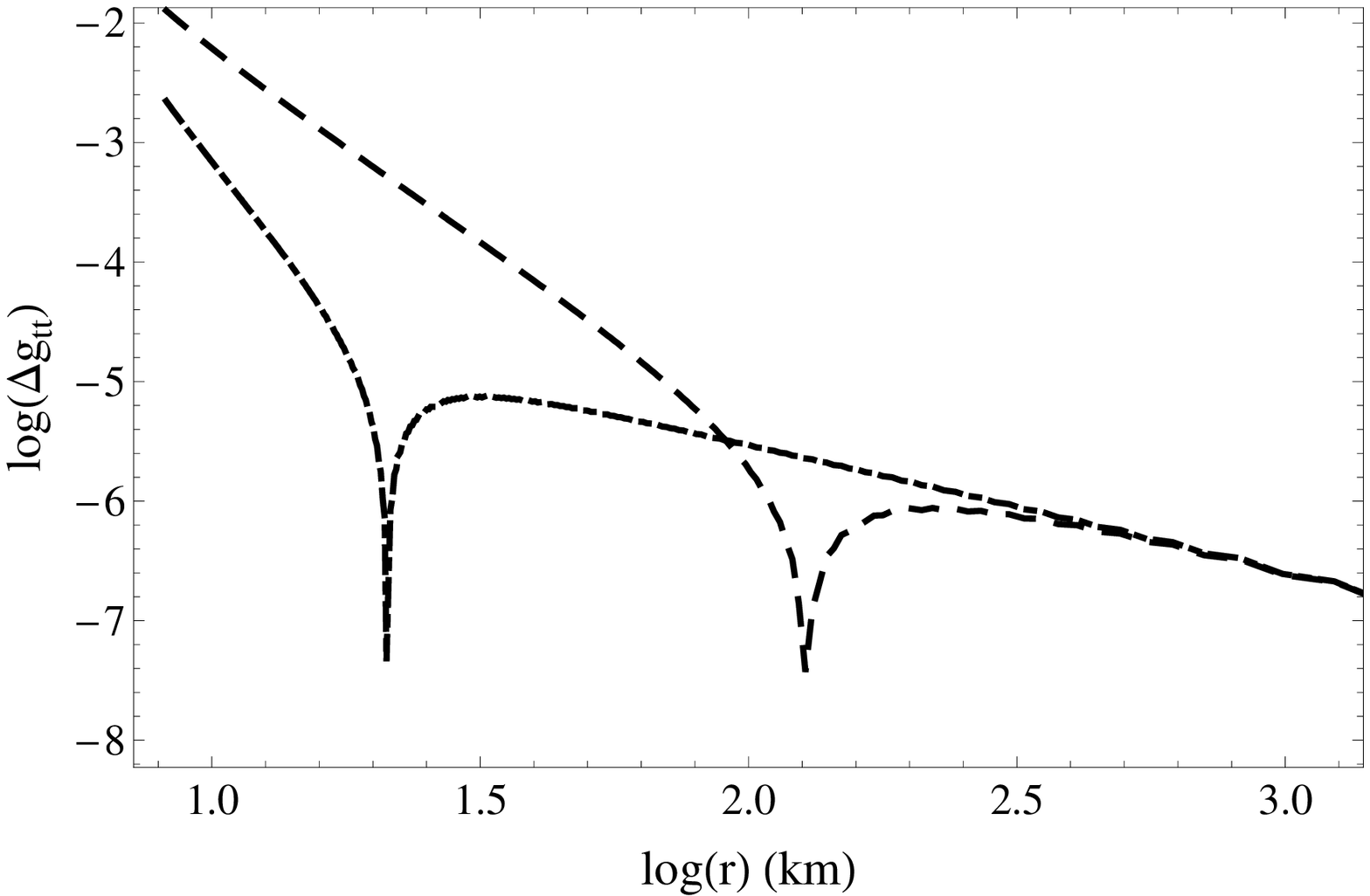}
  \includegraphics[width=0.32\textwidth]{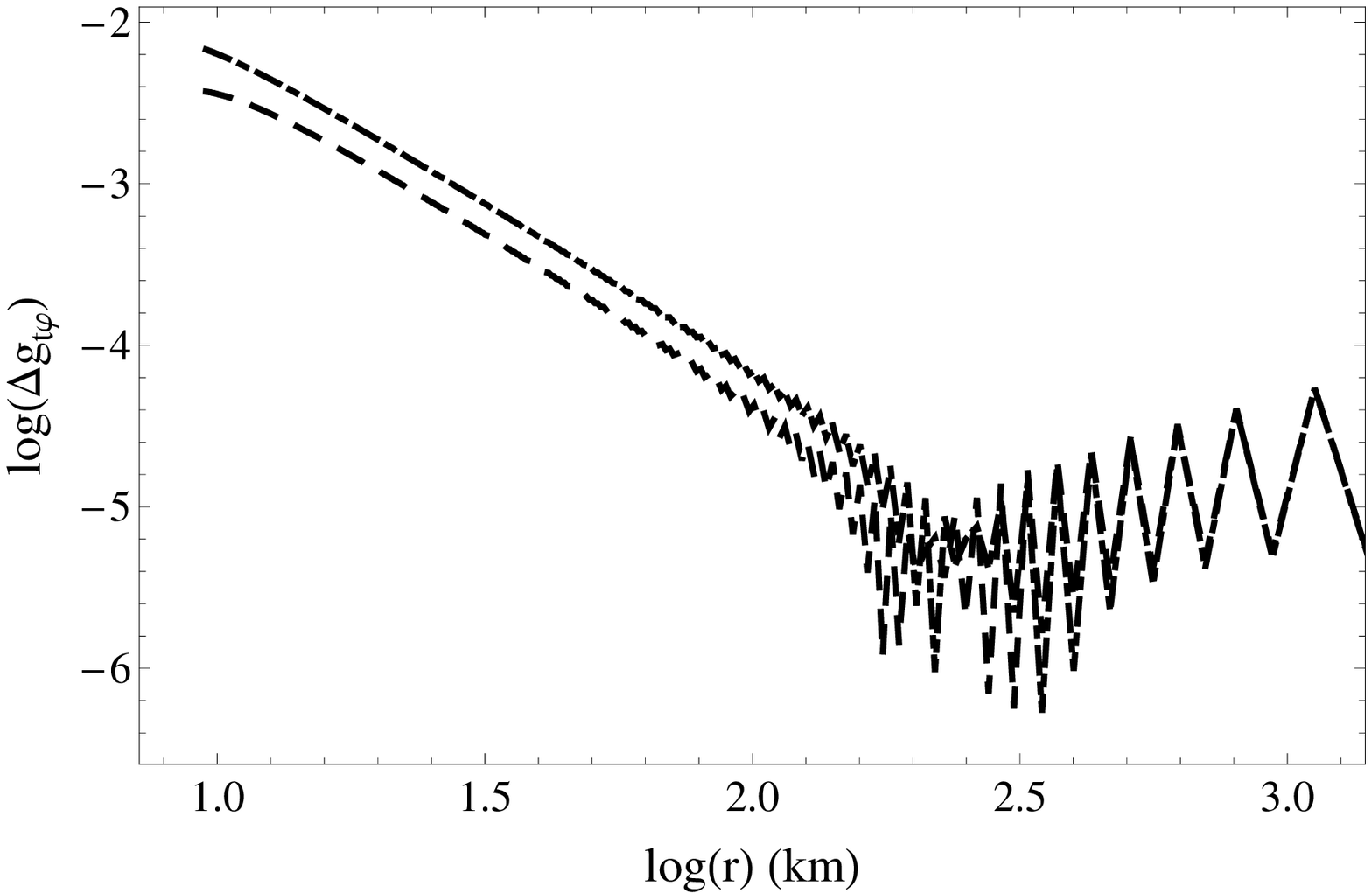}
  \includegraphics[width=0.32\textwidth]{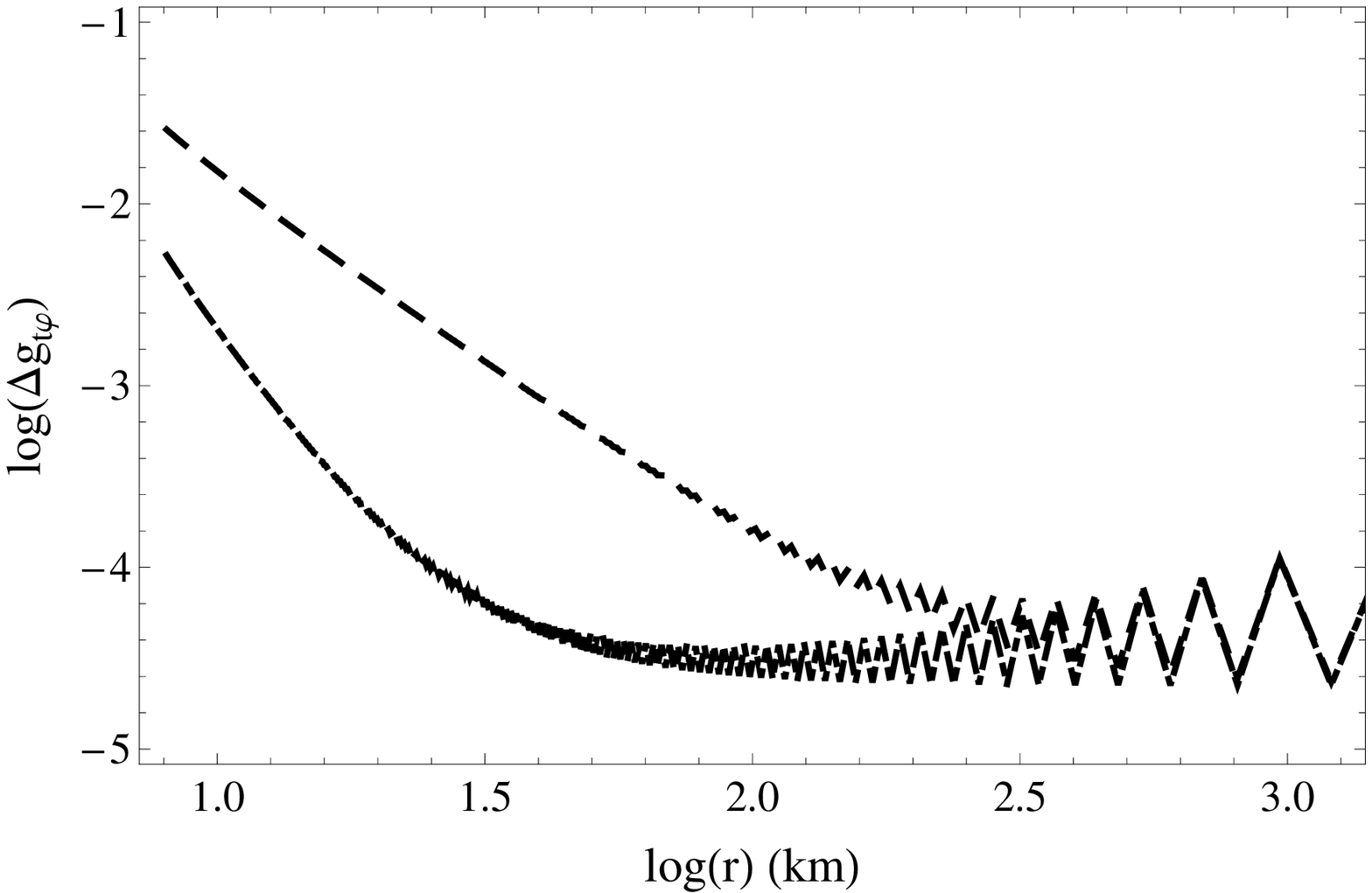}
  \includegraphics[width=0.32\textwidth]{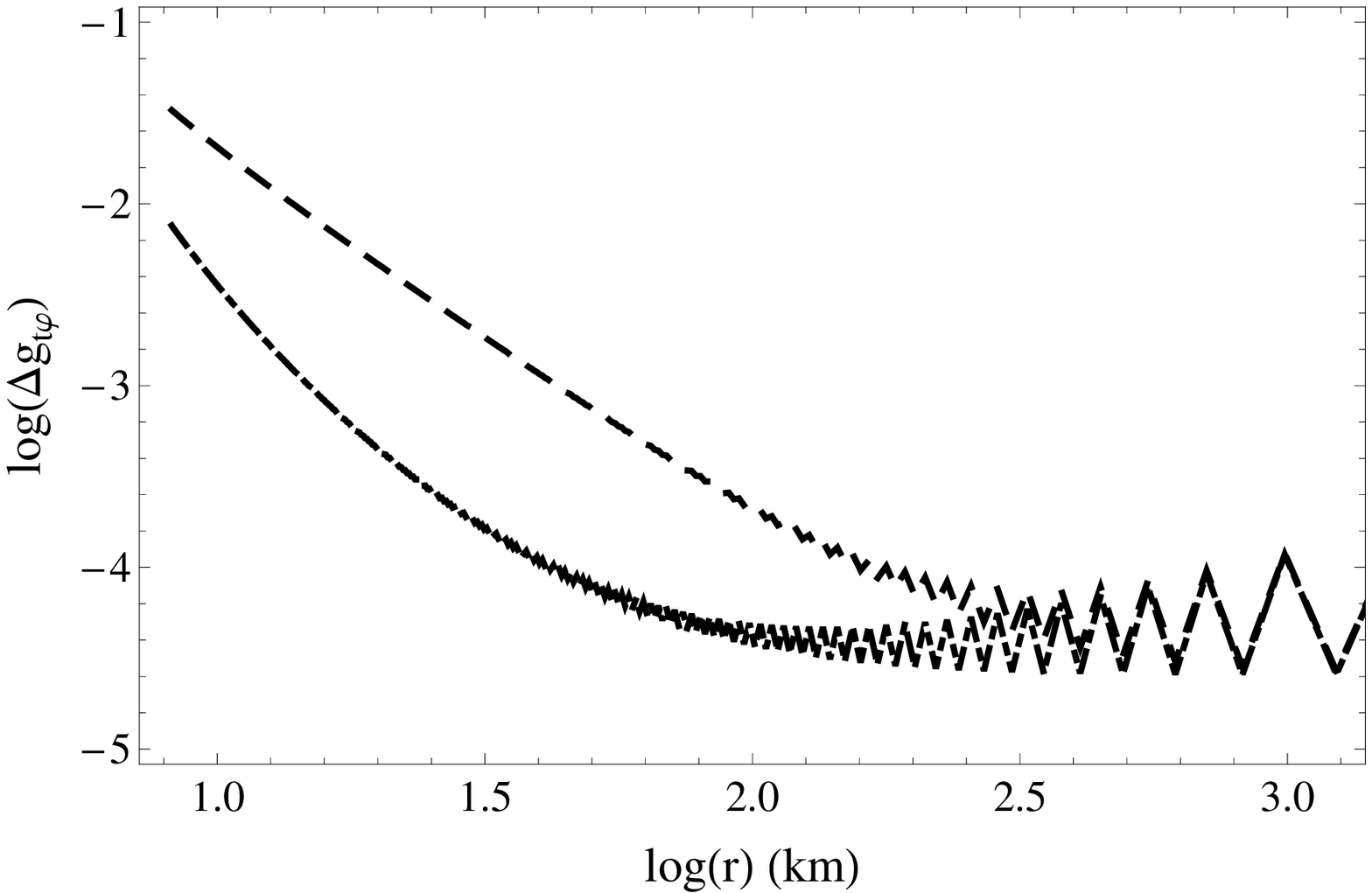}
  \caption{Plot of the comparison of the analytic against the numerical metric functions $g_{tt}$ (upper plots) and
  $g_{t\phi}$ (lower plots) for EOS AU. We show three typical models, one from each sequence.
  From left to right, the models are: $\#6$, $\#18$ and $\#29$. The dashed curves are the ones
  without the correction in the quadrupole and the dashed-dotted curves are the ones with the corrected quadrupole.}
  \protect\label{figmanko1}
\end{figure*}

\begin{figure*}

  \includegraphics[width=0.32\textwidth]{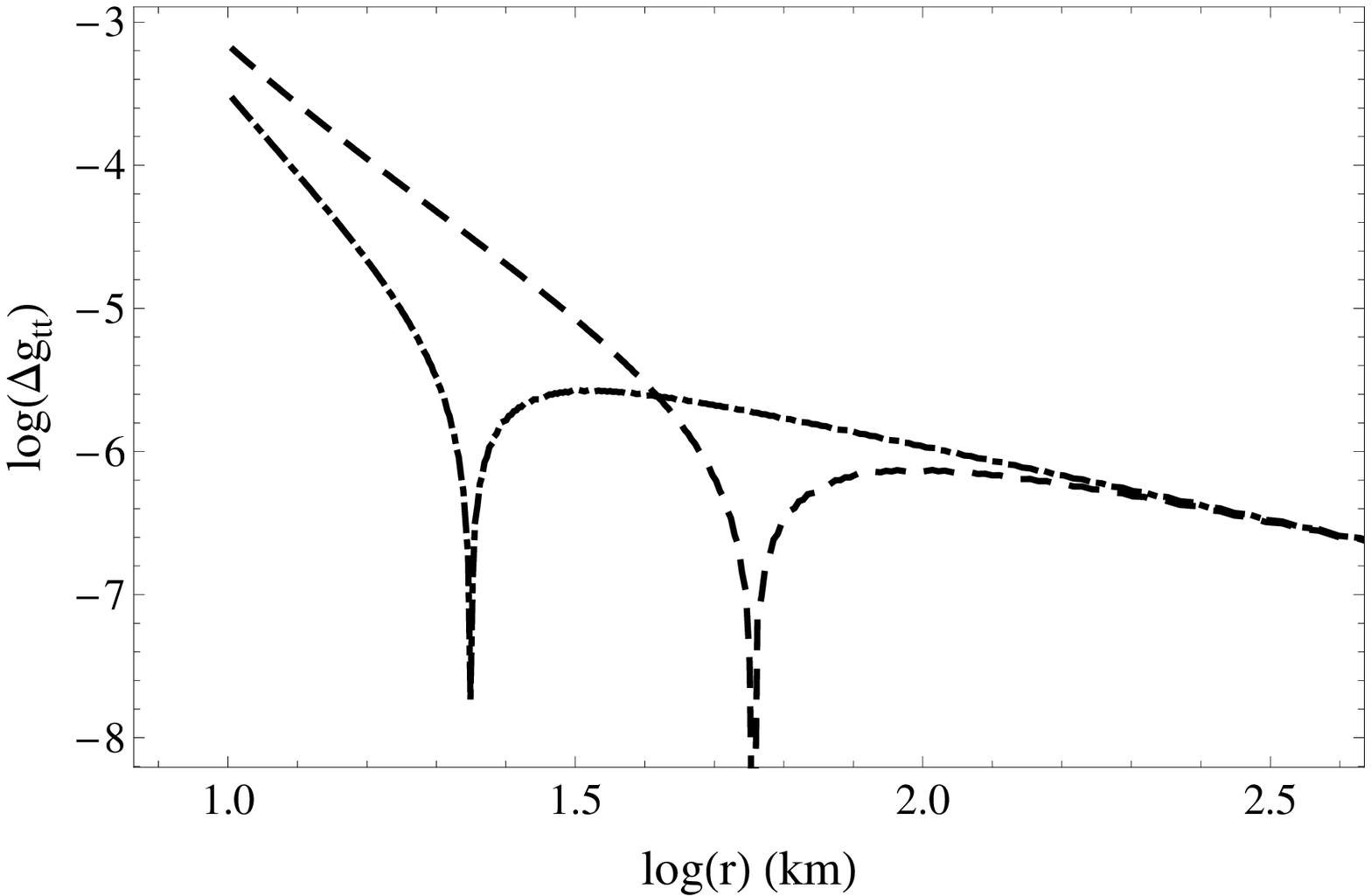}
  \includegraphics[width=0.32\textwidth]{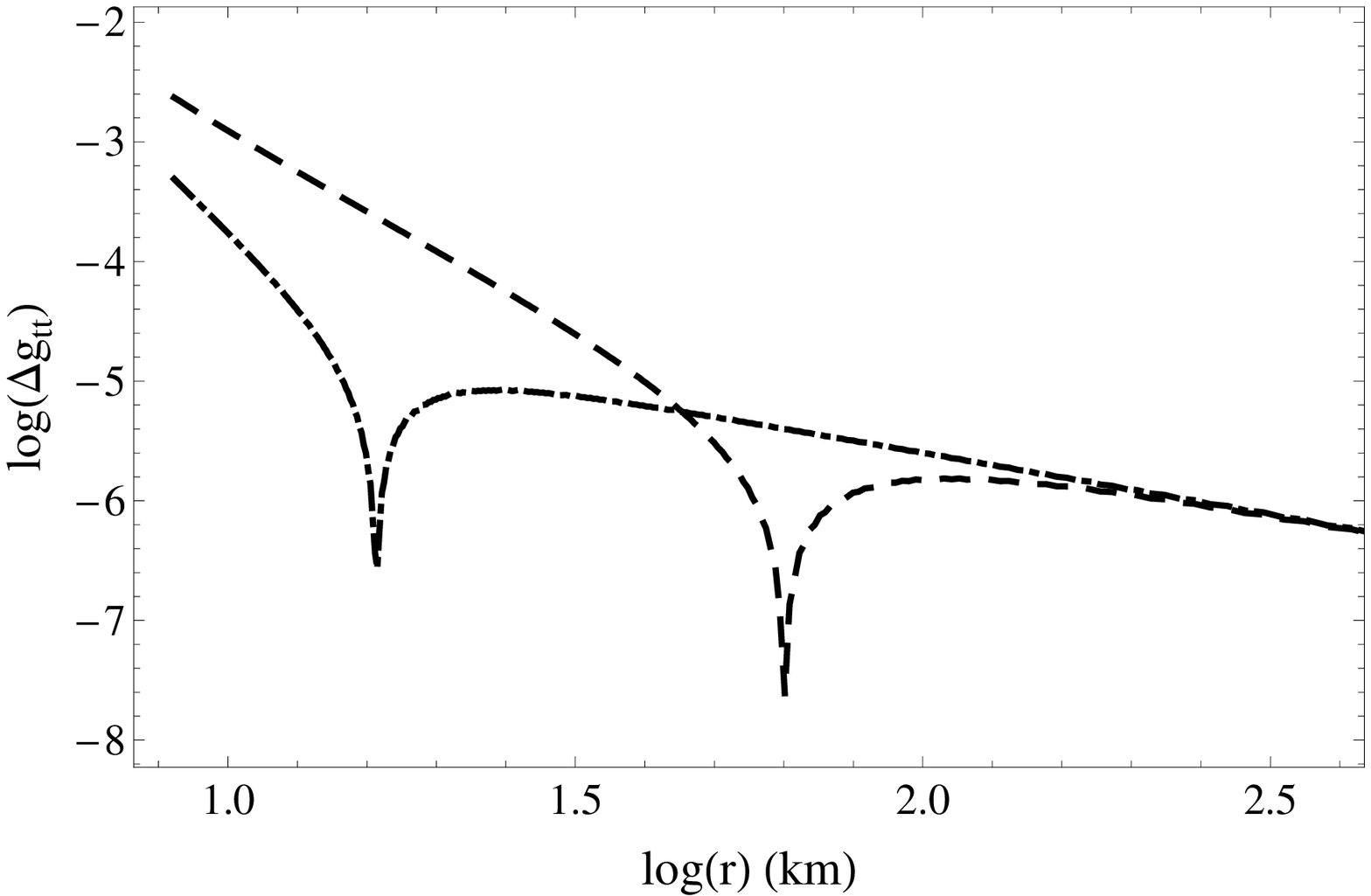}
  \includegraphics[width=0.32\textwidth]{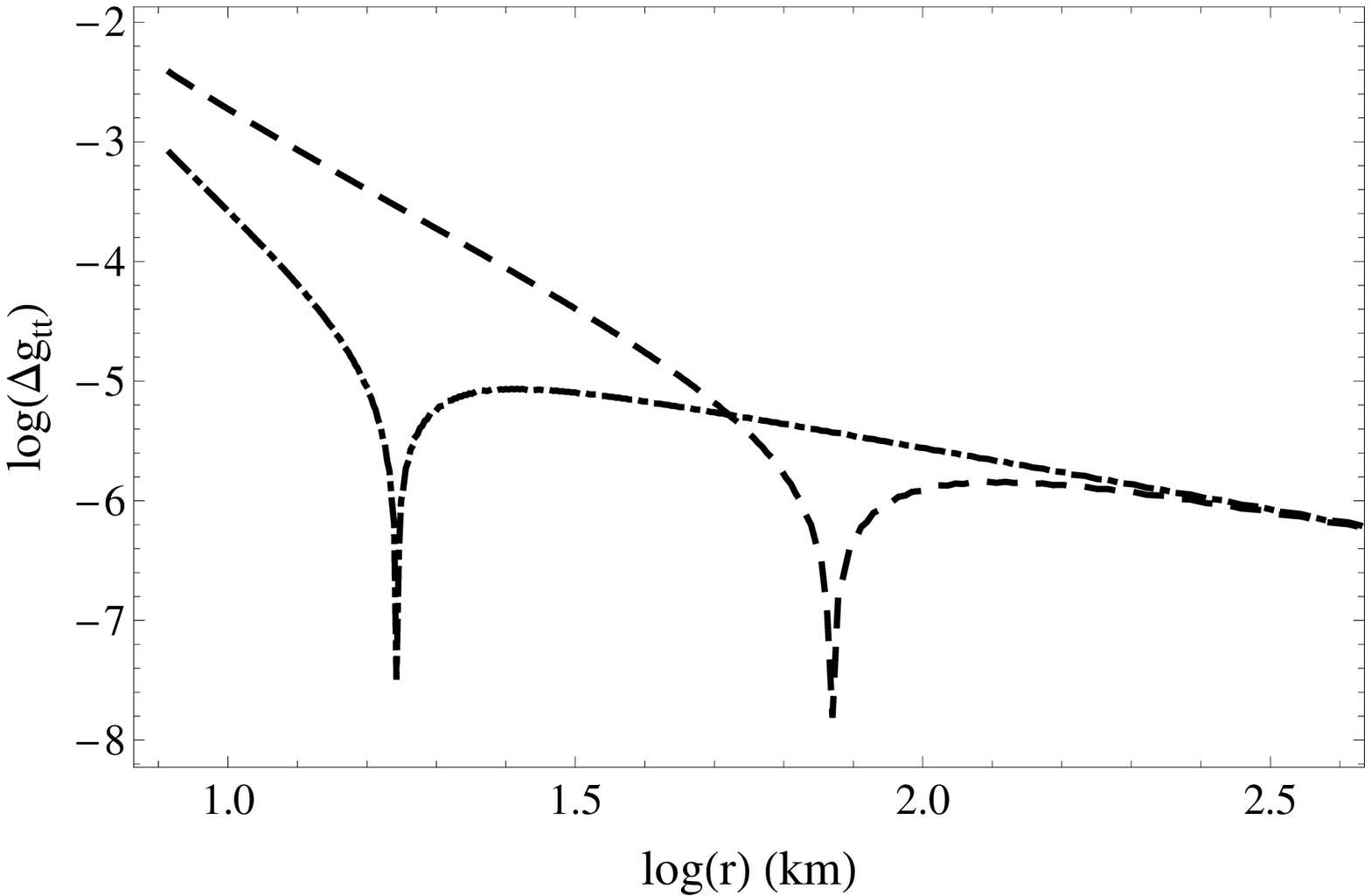}
  \includegraphics[width=0.32\textwidth]{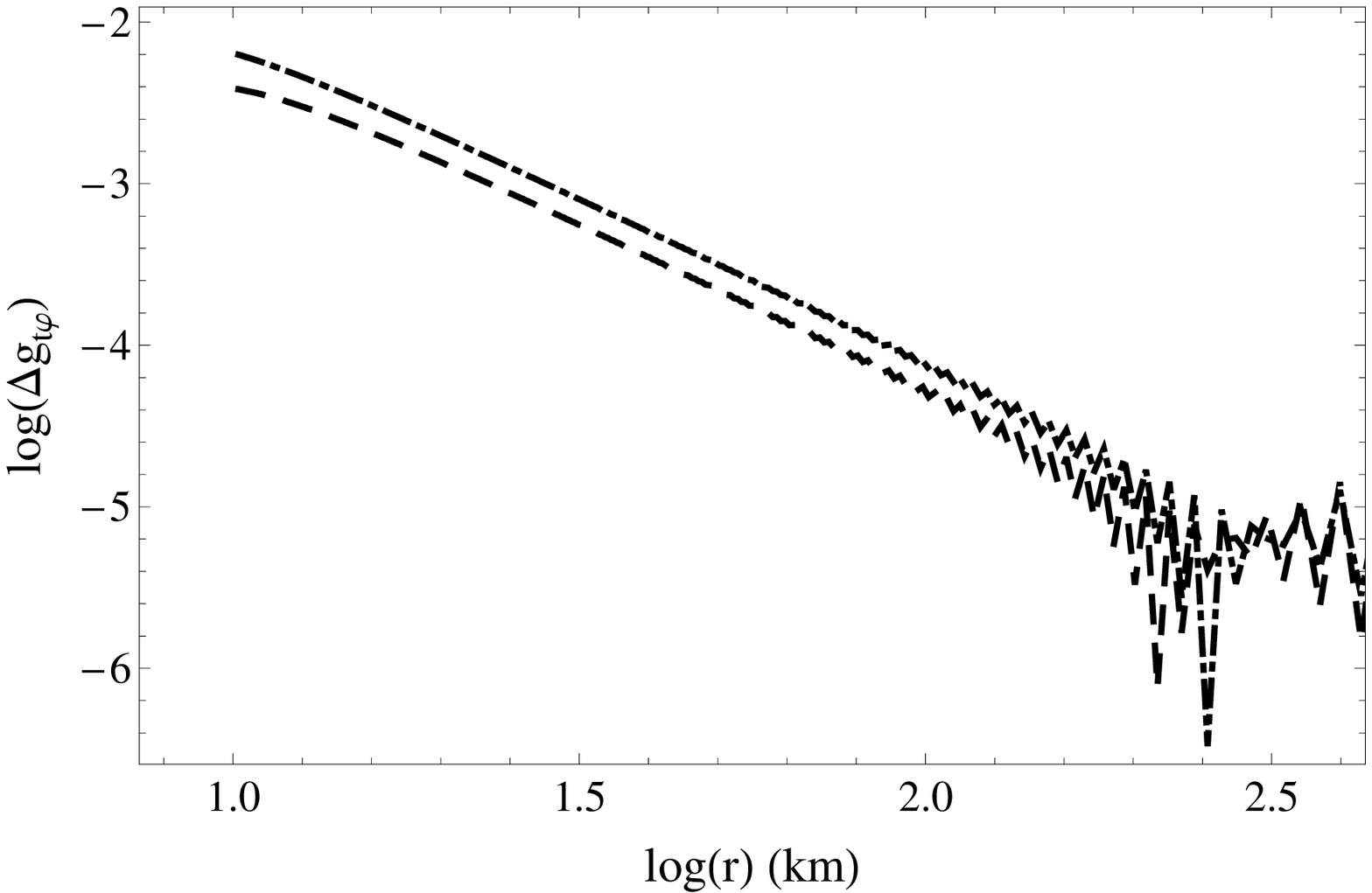}
  \includegraphics[width=0.32\textwidth]{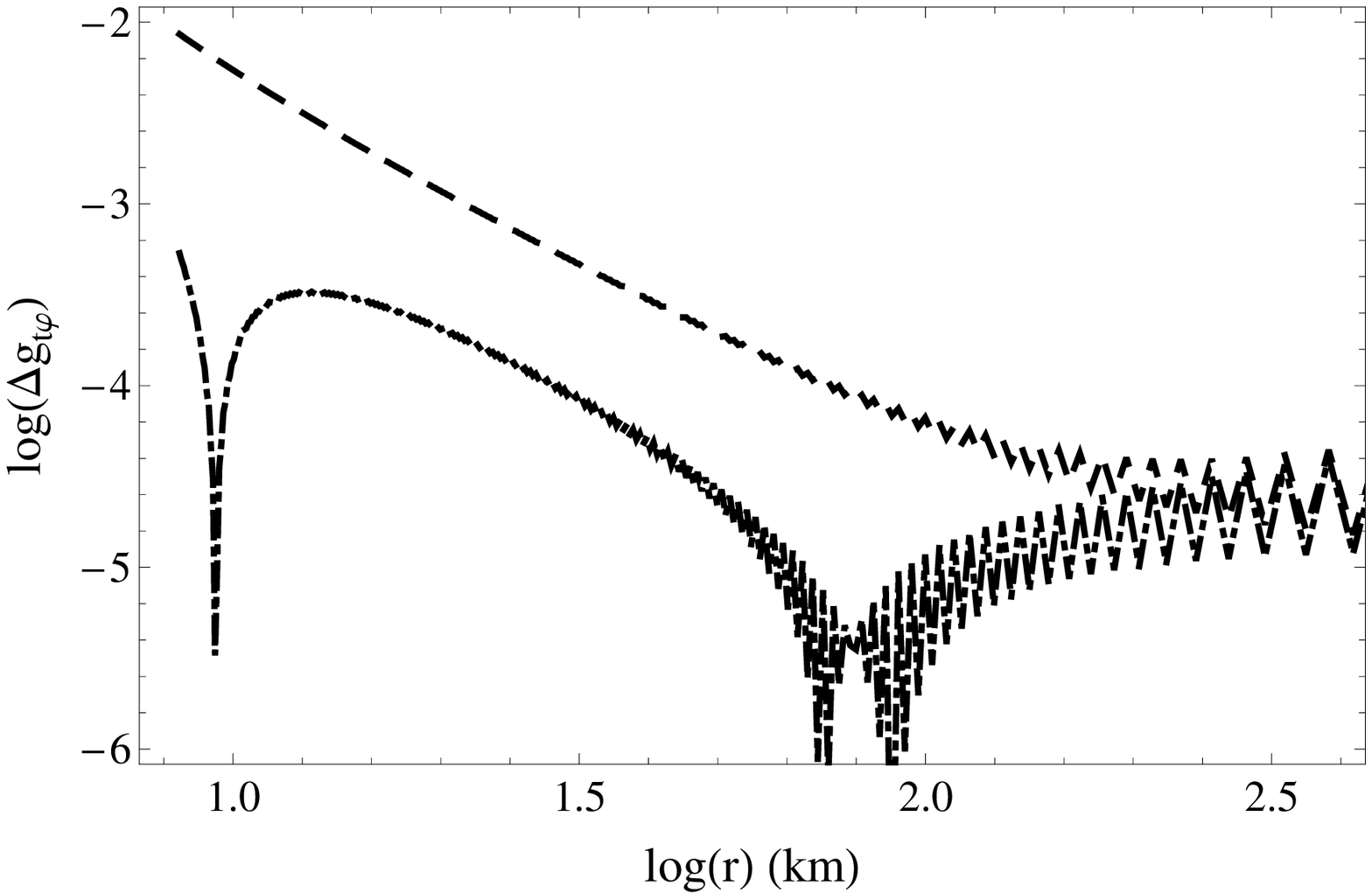}
  \includegraphics[width=0.32\textwidth]{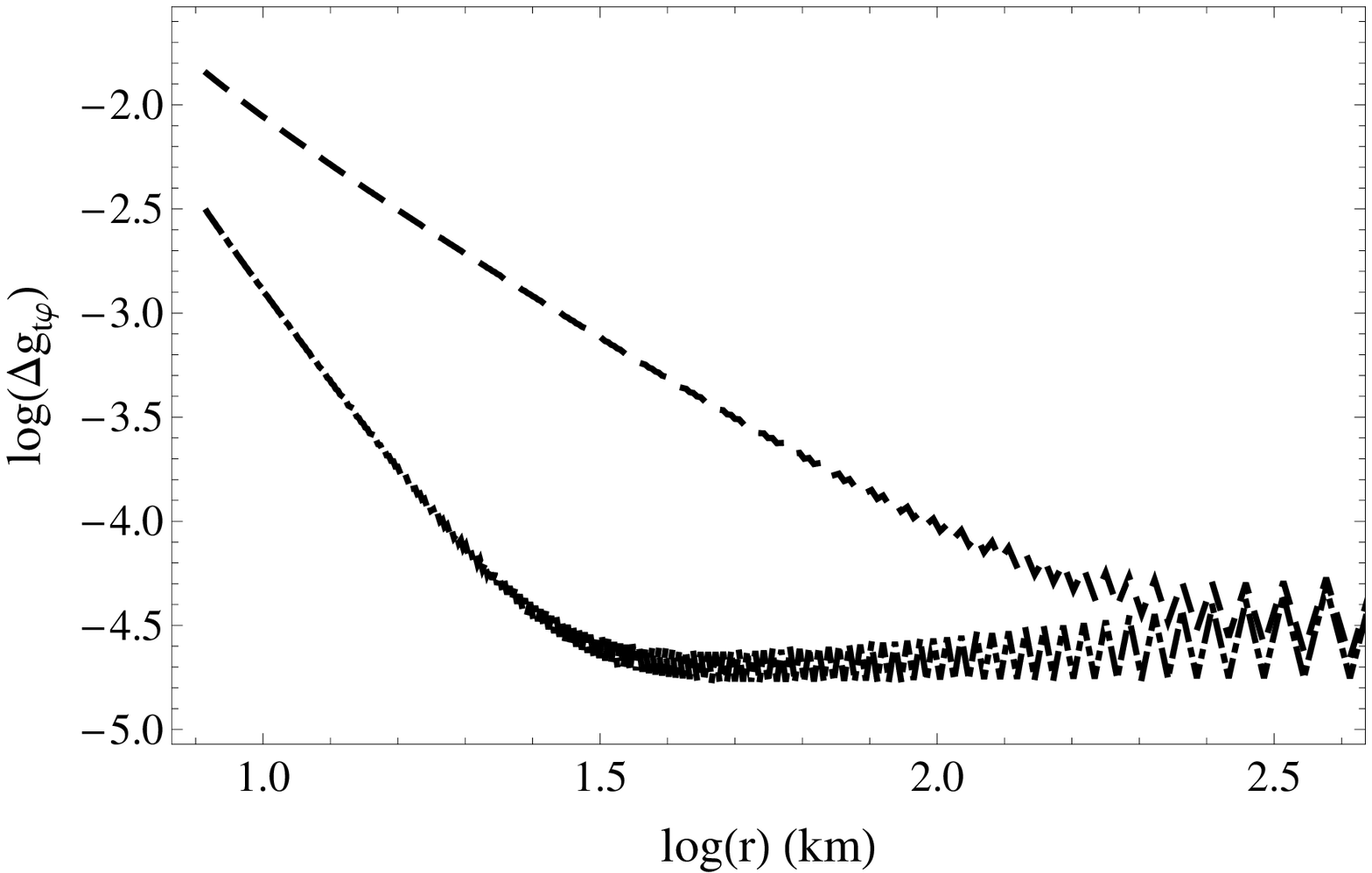}
  \caption{Same as Fig.~\ref{figmanko1} for EOS FPS.
  We show typical models from the three sequences of neutron stars.
  From left to right, the models are: $\#6$, $\#16$ and $\#27$.}
  \protect\label{figmanko2}
\end{figure*}

\begin{figure*}

  \includegraphics[width=0.32\textwidth]{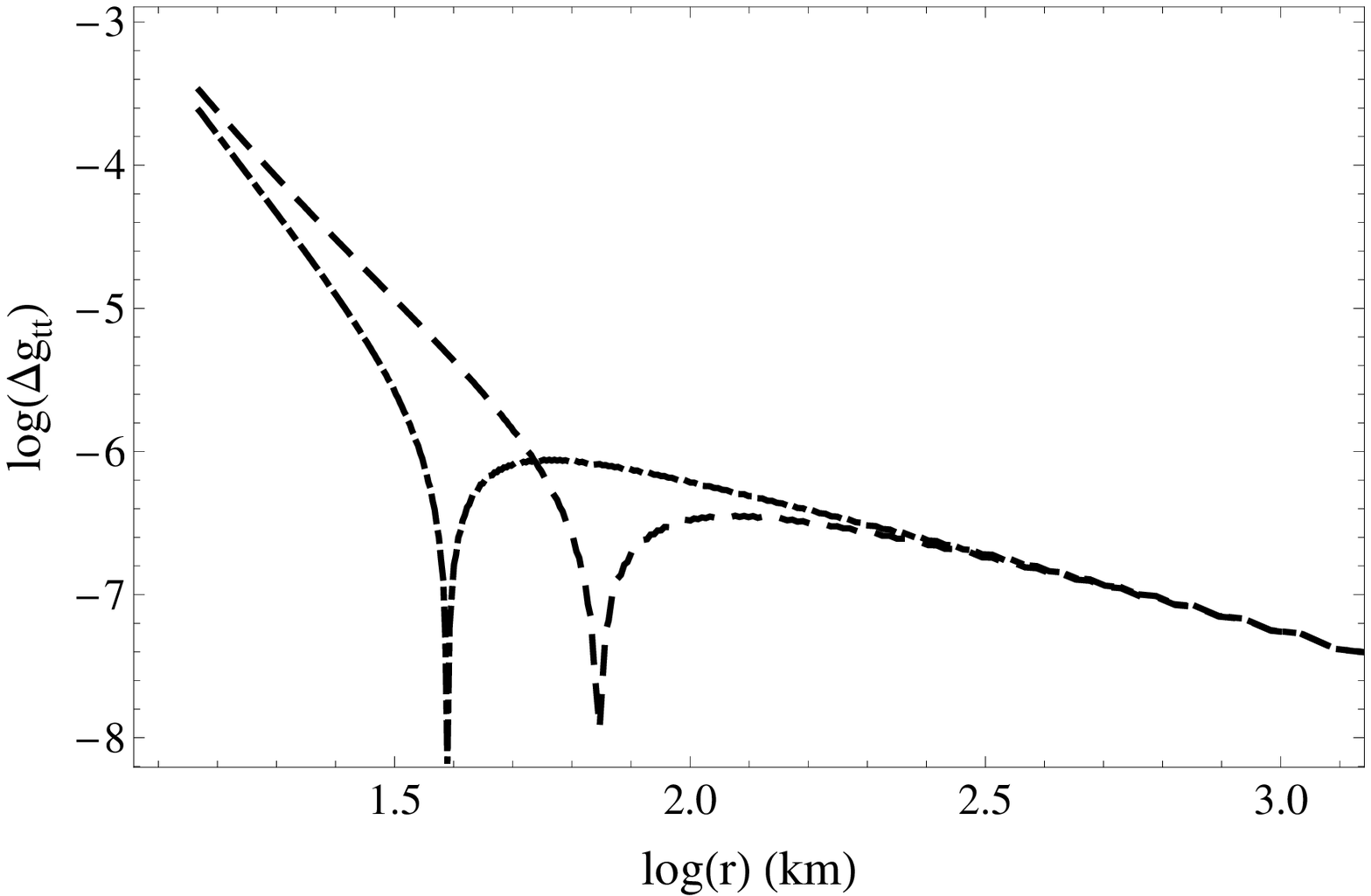}
  \includegraphics[width=0.32\textwidth]{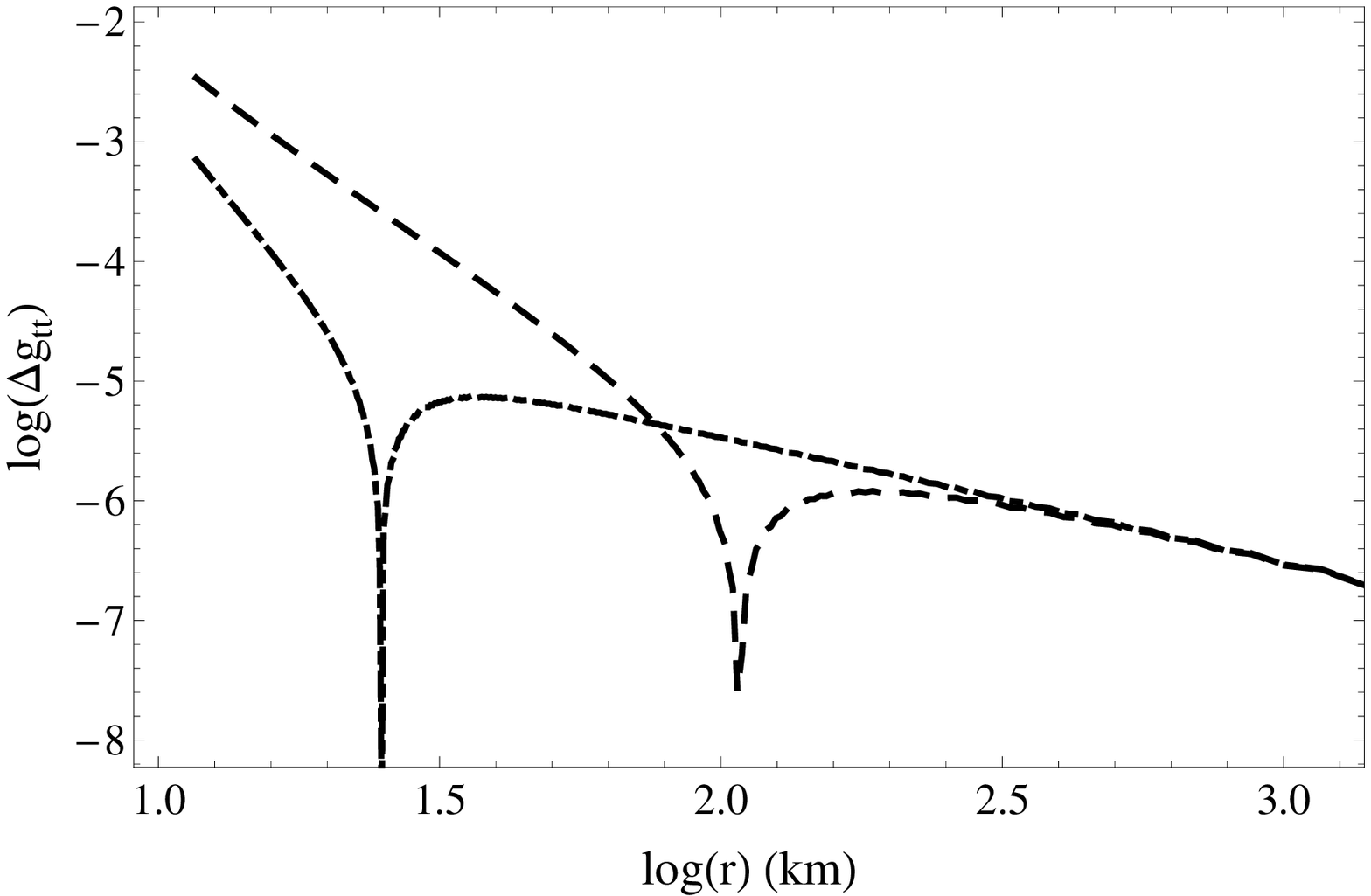}
  \includegraphics[width=0.32\textwidth]{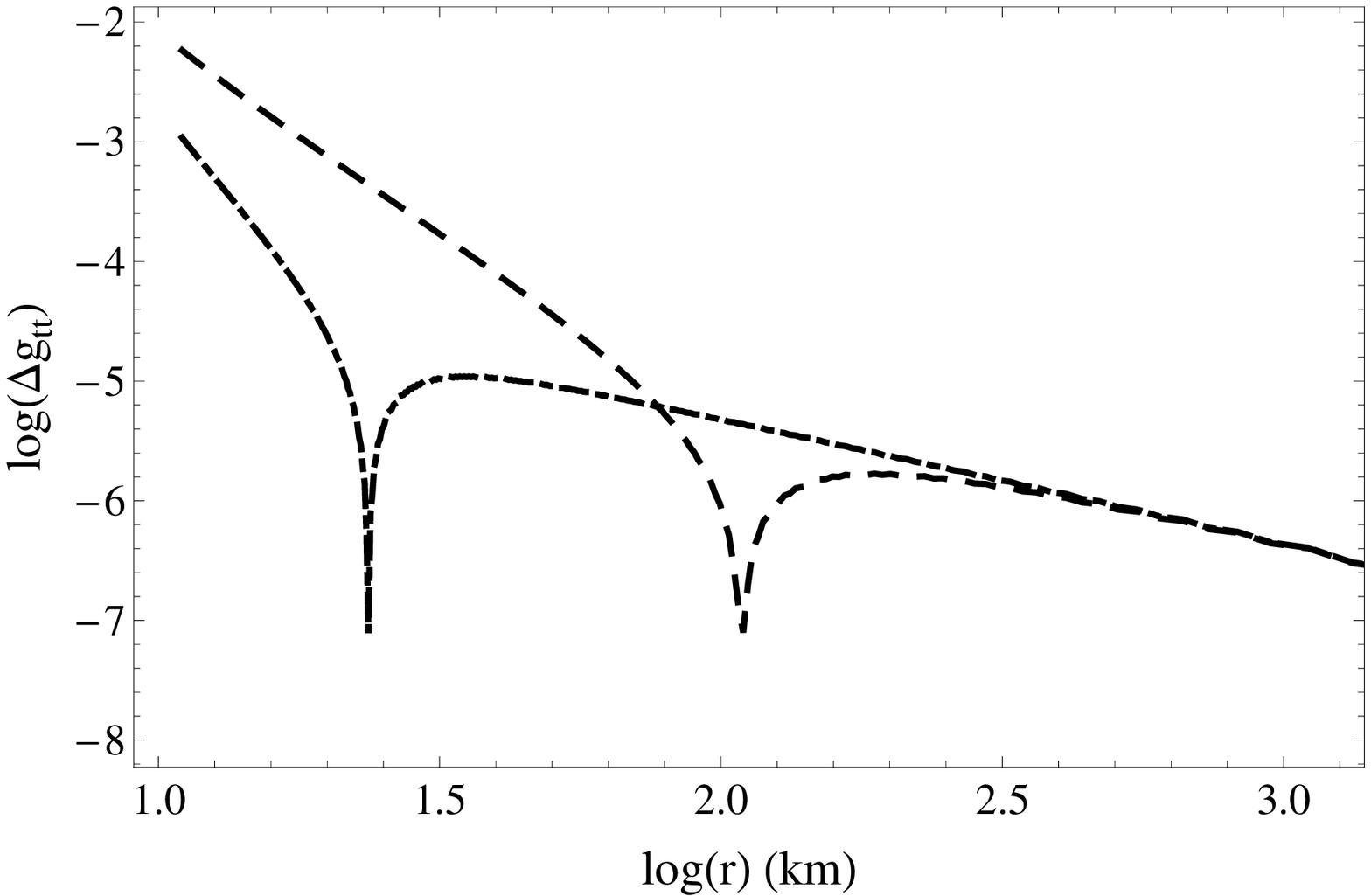}
  \includegraphics[width=0.32\textwidth]{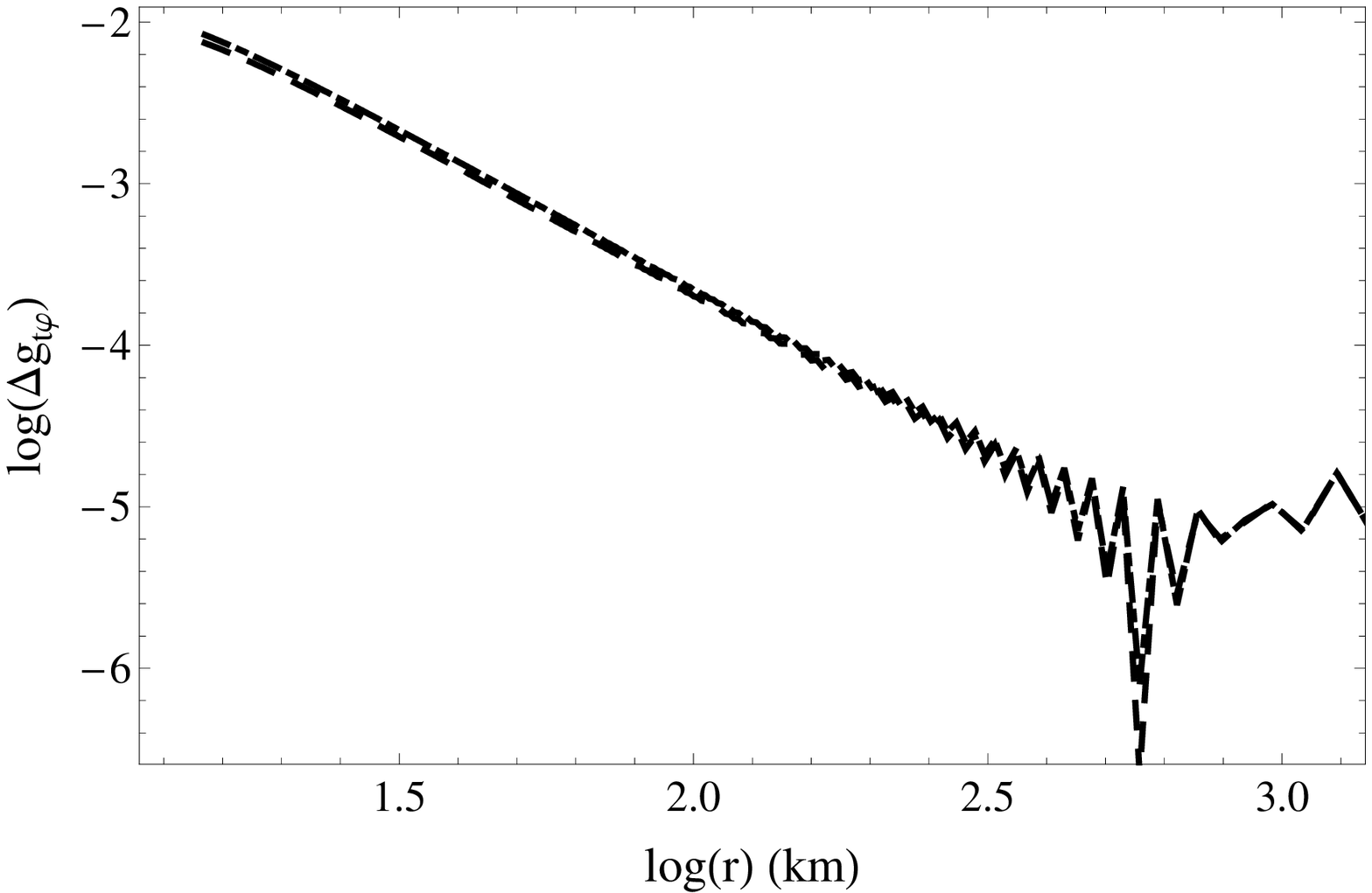}
  \includegraphics[width=0.32\textwidth]{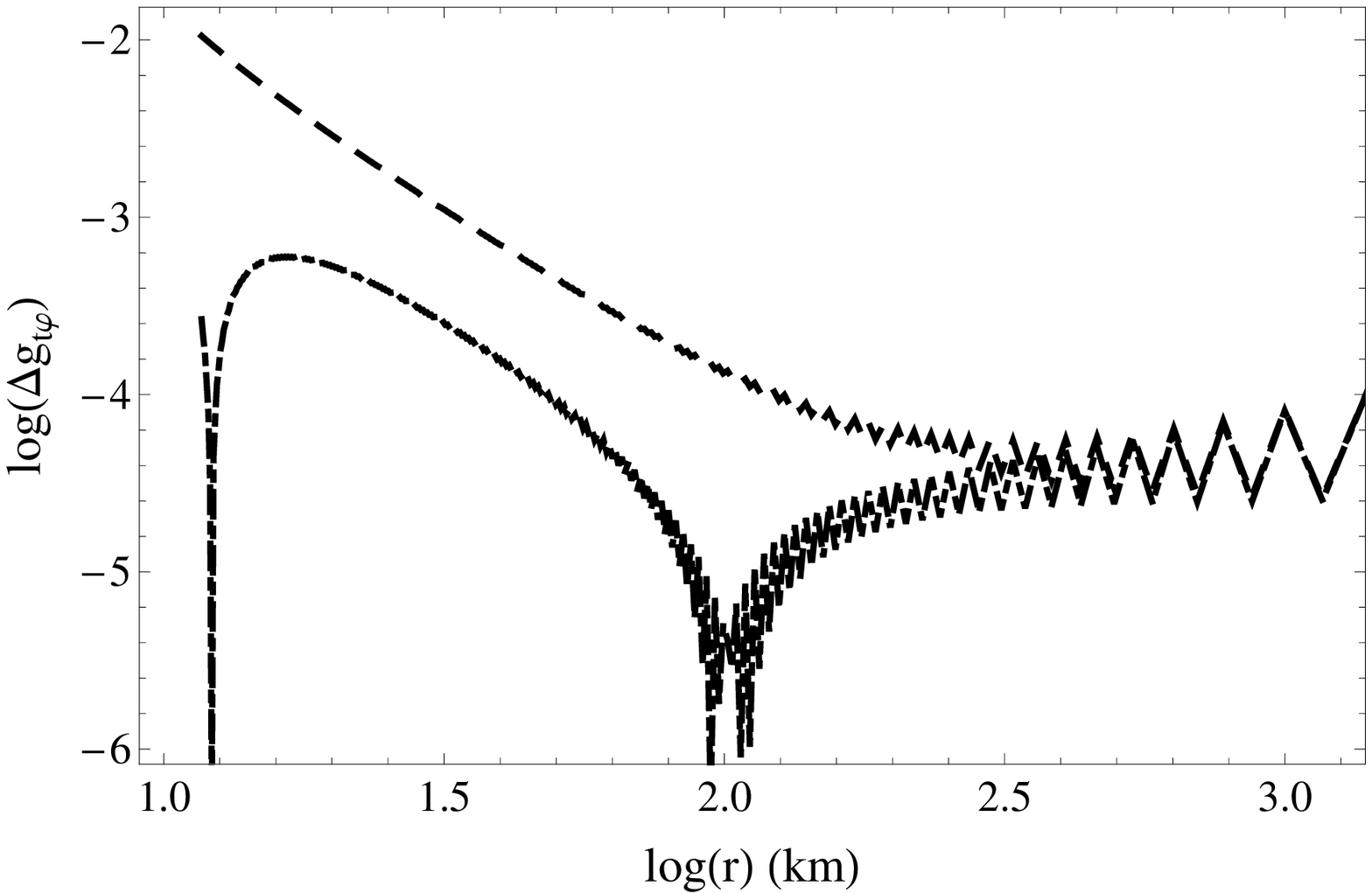}
  \includegraphics[width=0.32\textwidth]{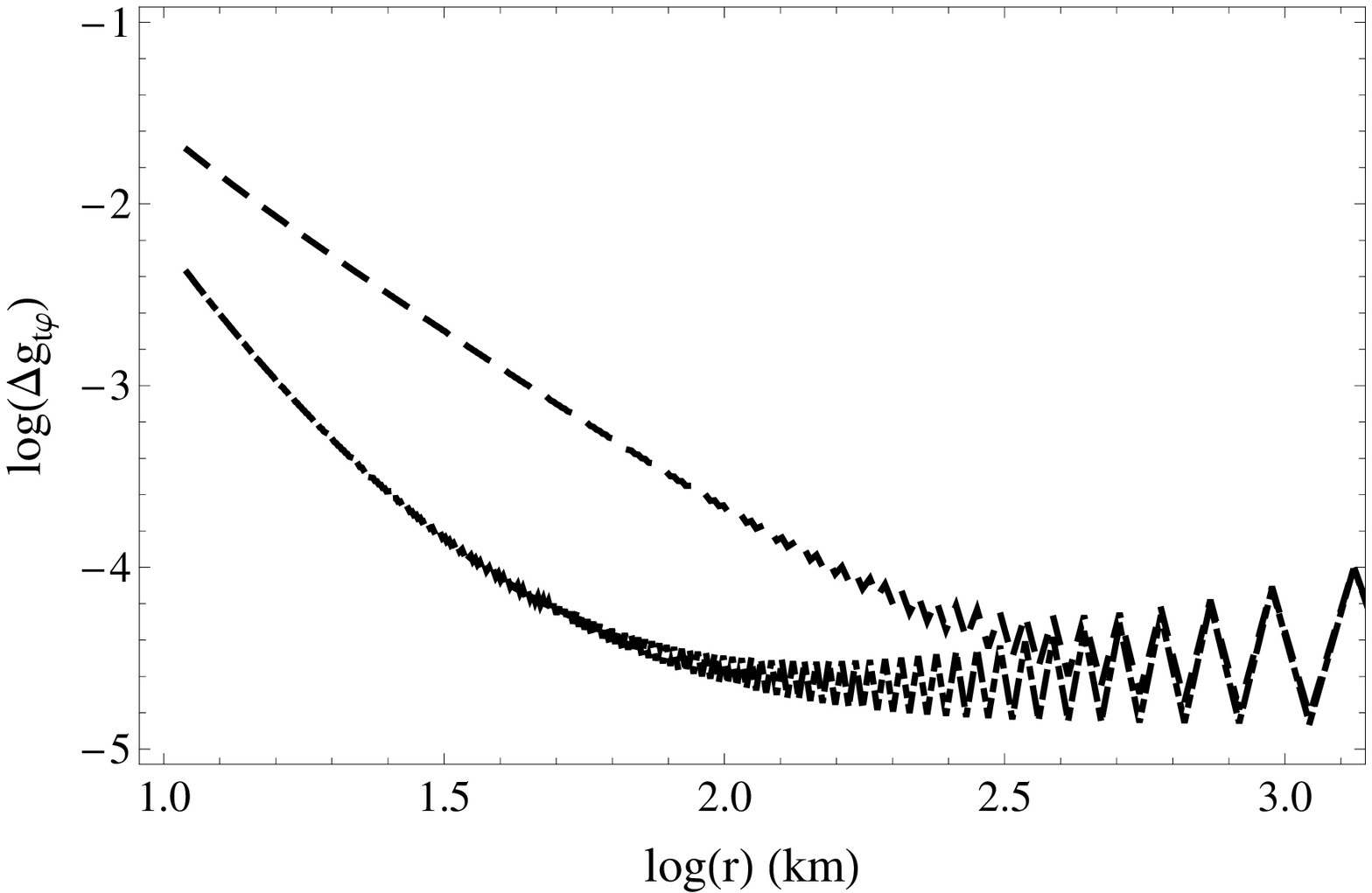}
  \caption{Same as Fig.~\ref{figmanko1} for EOS L.
  We show typical models from the three sequences of neutron stars.
  From left to right, the models are: $\#6$, $\#16$ and $\#26$.}
  \protect\label{figmanko3}
\end{figure*}

Finally in Figures \ref{figmanko1}, \ref{figmanko2},
\ref{figmanko3}, we have plotted for one model of each sequence
and each EOS the fractional difference between the actual
numerical metric functions and the corresponding Manko et al.
metric functions both when the analytic metrics were built with
the wrong and with the right quadrupole moment. These plots show
graphically the improvement caused by the correction of the
quadrupole moment. We should note that in the cases that the
correction in the quadrupole is small (which corresponds to the
first sequences of all EOSs), the mismatch between the analytic
and the numerical metric component $g_{t\phi}$ is slightly worse
after the correction. That effect is discussed in the main text.
As it can be seen though, the mismatch both before and after the
correction is relatively small in these cases.

\begin{table*}

 \centering
\begin{minipage}{0.5 \textwidth}
  \caption{EOS AU.}
 % \begin{scriptsize}
  \centering
   \begin{ruledtabular}
\begin{tabular}{c c c c c c c c}
 $\#$ & $M$ & $j$ & $M_2$ & $S_3$ & $b$ & $\Delta M_2$(\%) & $\Delta S_3$(\%) \\
 \hline
1 & 2.069 & 0.    & 0.    & 0.    & -0.25 &   -   &   -   \\ %& - & - & - & - & - & - \\
2 & 2.072 & 0.201 & -1.45 & -1.14 & -0.24 & 3.956 & 3.756 \\ %& - & - & - & - & - & - \\
3 & 2.078 & 0.312 & -3.47 & -4.28 & -0.23 & 3.993 & 3.779 \\ %& -4.92 & -4.56 & 0.00019 & 0.00004 & 0.00049 & 0.00115 \\
4 & 2.087 & 0.414 & -6.08 & -10.0 & -0.23 & 4.032 & 3.805 \\ %& -12.3 & -11.6 & 0.00037 & 0.00007 & 0.00263 & 0.00376 \\
5 & 2.092 & 0.474 & -7.99 & -15.1 & -0.22 & 4.038 & 3.799 \\ %& -18.2 & -17.2 & 0.00052 & 0.00014 & 0.00301 & 0.00447 \\
6 & 2.097 & 0.529 & -9.96 & -21.2 & -0.21 & 4.044 & 3.792 \\ %& -24.7 & -23.4 & 0.00067 & 0.00022 & 0.00260 & 0.00437 \\
7 & 2.103 & 0.578 & -11.9 & -27.9 & -0.21 & 4.041 & 3.775 \\ %& -31.5 & -30.0 & 0.00081 & 0.00032 & 0.00162 & 0.00360 \\
8 & 2.108 & 0.616 & -13.6 & -34.1 & -0.20 & 4.029 & 3.751 \\ %& -37.6 & -35.8 & 0.00090 & 0.00039 & 0.00069 & 0.00260 \\
9 & 2.111 & 0.650 & -15.2 & -40.5 & -0.20 & 4.008 & 3.716 \\ %& -43.5 & -41.5 & 0.00095 & 0.00044 & 0.00109 & 0.00149 \\
10 & 2.112 & 0.661 & -15.7 & -42.7 & -0.20 & 4.002 & 3.706 \\ %& -45.5 & -43.4 & 0.00095 & 0.00044 & 0.00140 & 0.00117 1 & \\
\hline
11 & 3.151 & 0.    &  0.   &  0.   & -0.25 &   -   &   -   \\ %& - & - & - & - & - & - \\
12 & 3.164 & 0.194 & -1.68 & -1.37 & -0.24 & 21.63 & 31.69 \\ %& - & - & - & - & - & - \\
13 & 3.183 & 0.309 & -4.46 & -5.97 & -0.23 & 20.41 & 28.99 \\ %& - & - & - & - & - & - \\
14 & 3.207 & 0.406 & -8.08 & -14.5 & -0.22 & 19.35 & 26.72 \\ %& - & - & - & - & - & - \\
15 & 3.231 & 0.485 & -12.0 & -26.6 & -0.20 & 18.43 & 24.82 \\ %& - & - & - & - & - & - \\
16 & 3.253 & 0.550 & -16.1 & -41.4 & -0.19 & 17.62 & 23.18 \\ %& - & - & - & - & - & - \\
17 & 3.273 & 0.603 & -20.2 & -58.1 & -0.18 & 16.92 & 21.80 \\ %& - & - & - & - & - & - \\
18 & 3.291 & 0.645 & -23.9 & -75.1 & -0.17 & 16.33 & 20.67 \\ %& -74.9 & -56.7 & 0.00529 & 0.00068 & 0.04020 & 0.00627 \\
19 & 3.304 & 0.676 & -27.0 & -90.4 & -0.17 & 15.86 & 19.77 \\ %& -90.2 & -69.8 & 0.00534 & 0.00078 & 0.04087 & 0.00728 \\
20 & 3.318 & 0.706 & -30.3 & -107. & -0.16 & 15.40 & 18.88 \\ %& -106. & -84.0 & 0.00486 & 0.00073 & 0.03790 & 0.00766 1 & \\
\hline
21 & 3.388 & 0.510 & -12.8 & -27.8 & -0.19 & 26.05 & 42.10 \\ %& - & - & - & - & - & - \\
22 & 3.388 & 0.510 & -12.9 & -28.1 & -0.19 & 25.75 & 41.32 \\ %& - & - & - & - & - & - \\
23 & 3.390 & 0.514 & -13.2 & -29.3 & -0.19 & 25.17 & 39.83 \\ %& - & - & - & - & - & - \\
24 & 3.393 & 0.520 & -13.7 & -31.1 & -0.19 & 24.60 & 38.38 \\ %& - & - & - & - & - & - \\
25 & 3.405 & 0.547 & -15.8 & -38.7 & -0.19 & 23.18 & 34.89 \\ %& - & - & - & - & - & - \\
26 & 3.422 & 0.587 & -19.1 & -51.6 & -0.18 & 21.76 & 31.57 \\ %& - & - & - & - & - & - \\
27 & 3.441 & 0.626 & -22.7 & -67.1 & -0.17 & 20.65 & 29.07 \\ %& - & - & - & - & - & - \\
28 & 3.458 & 0.659 & -26.0 & -82.7 & -0.17 & 19.79 & 27.20 \\ %& - & - & - & - & - & - \\
29 & 3.477 & 0.694 & -30.1 & -103. & -0.16 & 18.91 & 25.31 \\ %& -101. & -73.1 & 0.00705 & 0.00088 & 0.06104 & 0.01141 \\
30 & 3.487 & 0.713 & -32.5 & -115. & -0.16 & 18.45 & 24.34 \\ %& -113. & -83.7 & 0.00633 & 0.00078 & 0.05436 & 0.01051

\end{tabular}
 \end{ruledtabular}
 \protect\label{num1}
%\end{scriptsize}
\end{minipage}
\end{table*}

\begin{table*}
 \centering
\begin{minipage}{0.5 \textwidth}
  \caption{EOS AU.}
 % \begin{small}
  \centering
 \begin{ruledtabular}
\begin{tabular}{ c c c c c c}
% $M$ & $j$ & $\Delta Q$(\%) & $\Delta S_3$(\%)
$\#$  & $\Delta S_3^M$(\%) &
 $\frac{\sigma_{tt}|_{old}}{\sigma_{tt}|_{new}}$ & $\sigma_{tt}|_{new}$ &
 $\frac{\sigma_{t\phi}|_{old}}{\sigma_{t\phi}|_{new}}$ & $\sigma_{t\phi}|_{new}$
 \\ \hline
% 1 & -     & -     & -       & -       & - \\
% 2 & -     & -     & -       & -       & - \\
 3 & 7.319 & 4.760 & 0.00004 & 0.42877 & 0.00115 \\
 4 & 5.474 & 5.150 & 0.00007 & 0.70046 & 0.00376 \\
 5 & 5.159 & 3.690 & 0.00014 & 0.67202 & 0.00447 \\
 6 & 4.981 & 2.928 & 0.00022 & 0.59514 & 0.00437 \\
 7 & 4.857 & 2.506 & 0.00032 & 0.44967 & 0.00360 \\
 8 & 4.770 & 2.294 & 0.00039 & 0.26800 & 0.00260 \\
 9 & 4.690 & 2.169 & 0.00044 & 0.73064 & 0.00149 \\
10 & 4.668 & 2.150 & 0.00044 & 1.20269 & 0.00117
 \\ \hline
18 & 24.26 & 7.692 & 0.00068 & 6.4051 & 0.00627 \\
19 & 22.60 & 6.834 & 0.00078 & 5.61239 & 0.00728 \\
20 & 21.32 & 6.577 & 0.00073 & 4.94567 & 0.00766
 \\ \hline
29 & 27.99 & 7.985 & 0.00088 & 5.34521 & 0.01141 \\
30 & 26.49 & 8.058 & 0.00078 & 5.16811 & 0.01051

\end{tabular}
 \end{ruledtabular}
 \protect\label{anal1}
%\end{small}
\end{minipage}
\end{table*}

%%%%%%%%%%%%%%%%%%%%%%%%%%%%%%%%%%%%%%%%%%%%%%%%%%%%%%%%%%%%%%%%%%%%%%%%%%%%%%%%%%%%%%5

\begin{table*}

 \centering
\begin{minipage}{0.5 \textwidth}
  \caption{EOS FPS.}
%  \begin{scriptsize}
  \centering
   \begin{ruledtabular}
\begin{tabular}{c c c c c c c c}
 $\#$ & $M$ & $j$ & $M_2$ & $S_3$ & $b$ & $\Delta M_2$(\%) & $\Delta S_3$(\%) \\
 \hline
1  & 2.067 & 0. & 0 & 0 & -0.25 & - & - \\ %& - & - & - & - & - & - \\
2  & 2.071 & 0.201 & -1.54 & -1.25 & -0.24 & 3.691 & 3.404 \\ %& - & - & - & - & - & - \\
3  & 2.077 & 0.301 & -3.45 & -4.22 & -0.23 & 3.685 & 3.385 \\ %& -4.85 & -4.52 & 0.00015 & 0.00003 & 0.00045 & 0.00096 \\
4  & 2.083 & 0.385 & -5.64 & -8.88 & -0.23 & 3.667 & 3.355 \\ %& -10.9 & -10.4 & 0.00027 & 0.00005 & 0.00216 & 0.00297 \\
5  & 2.087 & 0.452 & -7.85 & -14.6 & -0.22 & 3.634 & 3.311 \\ %& -17.6 & -16.8 & 0.00041 & 0.00012 & 0.00270 & 0.00380 \\
6  & 2.093 & 0.507 & -9.94 & -20.9 & -0.22 & 3.610 & 3.275 \\ %& -24.4 & -23.3 & 0.00054 & 0.00020 & 0.00245 & 0.00377 \\
7  & 2.098 & 0.557 & -12.0 & -28.0 & -0.21 & 3.575 & 3.228 \\ %& -31.7 & -30.4 & 0.00066 & 0.00028 & 0.00163 & 0.00311 \\
8  & 2.102 & 0.603 & -14.2 & -36.1 & -0.21 & 3.532 & 3.173 \\ %& -39.5 & -37.9 & 0.00075 & 0.00036 & 0.00061 & 0.00201 \\
9  & 2.106 & 0.636 & -15.9 & -42.9 & -0.20 & 3.504 & 3.136 \\ %& -45.8 & -44.0 & 0.00078 & 0.00039 & 0.00092 & 0.00108 \\
10  & 2.109 & 0.666 & -17.6 & -49.9 & -0.20 & 3.469 & 3.091 \\ %&-52.2 & -50.1 & 0.00069 & 0.00033 & 0.00148 & 0.00052 1  & \\
\hline
11  & 2.658 & 0. & 0 & 0 & -0.25 & - & - \\ %& - & - & - & - & - & - \\
12  & 2.664 & 0.163 & -0.94 & -0.64 & -0.24 & 13.74 & 16.07 \\ %& - & - & - & - & - & - \\
13  & 2.674 & 0.260 & -2.51 & -2.81 & -0.23 & 12.65 & 14.41 \\ %& - & - & - & - & - & - \\
14  & 2.686 & 0.349 & -4.81 & -7.43 & -0.23 & 11.63 & 12.90 \\ %& - & - & - & - & - & - \\
15  & 2.701 & 0.436 & -8.00 & -15.8 & -0.22 & 10.69 & 11.56 \\ %& - & - & - & - & - & - \\
16  & 2.714 & 0.500 & -11.0 & -25.6 & -0.21 & 10.05 & 10.65 \\ %& -26.4 & -22.3 & 0.00159 & 0.00024 & 0.00716 & 0.00047 \\
17  & 2.727 & 0.562 & -14.6 & -39.0 & -0.20 & 9.428 & 9.805 \\ %& -40.6 & -35.3 & 0.00186 & 0.00037 & 0.00782 & 0.00085 \\
18  & 2.736 & 0.602 & -17.4 & -50.4 & -0.19 & 9.026 & 9.261 \\ %& -52.2 & -45.9 & 0.00198 & 0.00047 & 0.00907 & 0.00116 \\
19  & 2.744 & 0.633 & -19.7 & -60.8 & -0.19 & 8.734 & 8.866 \\ %& -62.3 & -55.3 & 0.00201 & 0.00053 & 0.01010 & 0.00203 \\
20  & 2.750 & 0.654 & -21.4 & -69.0 & -0.18 & 8.543 & 8.607 \\ %& -70.1 & -62.4 & 0.00192 & 0.00053 & 0.01041 & 0.00266 1  & \\
\hline
21  & 2.823 & 0.427 & -6.54 & -11.7 & -0.21 & 17.55 & 22.14 \\ %& - & - & - & - & - & - \\
22  & 2.823 & 0.428 & -6.69 & -12.1 & -0.21 & 16.97 & 21.12 \\ %& - & - & - & - & - & - \\
23  & 2.825 & 0.432 & -6.98 & -12.9 & -0.21 & 16.40 & 20.16 \\ %& - & - & - & - & - & - \\
24  & 2.826 & 0.439 & -7.35 & -13.9 & -0.21 & 15.84 & 19.22 \\ %& - & - & - & - & - & - \\
25  & 2.829 & 0.450 & -7.88 & -15.4 & -0.21 & 15.30 & 18.34 \\ %& - & - & - & - & - & - \\
26  & 2.840 & 0.492 & -10.1 & -22.5 & -0.20 & 13.76 & 15.88 \\ %& - & - & - & - & - & - \\
27  & 2.856 & 0.552 & -13.8 & -35.7 & -0.20 & 12.35 & 13.73 \\ %& -35.8 & -29.1 & 0.00245 & 0.00038 & 0.01448 & 0.00238 \\
28  & 2.871 & 0.609 & -17.9 & -52.9 & -0.19 & 11.25 & 12.13 \\ %& -53.5 & -45.2 & 0.00258 & 0.00052 & 0.01500 & 0.00288 \\
29  & 2.882 & 0.647 & -21.1 & -67.6 & -0.18 & 10.64 & 11.25 \\ %& -67.9 & -58.3 & 0.00249 & 0.00057 & 0.01540 & 0.00392 \\
30  & 2.884 & 0.658 & -22.2 & -72.8 & -0.18 & 10.42 & 10.94 \\ %& -72.9 & -62.9 & 0.00232 & 0.00053 & 0.01459 & 0.00392

\end{tabular}
 \end{ruledtabular}
%\end{scriptsize}
\protect\label{num2}
\end{minipage}
\end{table*}

\begin{table*}
 \centering
\begin{minipage}{0.5 \textwidth}
  \caption{EOS FPS.}
 % \begin{scriptsize}
  \centering
 \begin{ruledtabular}
\begin{tabular}{ c c c c c c}
% $M$ & $j$ & $\Delta Q$(\%) & $\Delta S_3$(\%)
$\#$  & $\Delta S_3^M$(\%) &
 $\frac{\sigma_{tt}|_{old}}{\sigma_{tt}|_{new}}$ & $\sigma_{tt}|_{new}$ &
 $\frac{\sigma_{t\phi}|_{old}}{\sigma_{t\phi}|_{new}}$ & $\sigma_{t\phi}|_{new}$
 \\ \hline
% 2.067 & 0. & - & - & - & 1 & - & 1 & - \\
% 2.071 & 0.201 & 3.691 & 3.404 & - & 1 & - & 1 & - \\
% 2.077 & 0.301 & 3.685 & 3.385
3 & 6.730 & 4.062 & 0.00003 & 0.46717 & 0.00096 \\
% 2.083 & 0.385 & 3.667 & 3.355
4 & 5.056 & 4.805 & 0.00005 & 0.72818 & 0.00297 \\
% 2.087 & 0.452 & 3.634 & 3.311
5 & 4.633 & 3.397 & 0.00012 & 0.71192 & 0.00380 \\
% 2.093 & 0.507 & 3.610 & 3.275
6 & 4.423 & 2.699 & 0.00020 & 0.64891 & 0.00377 \\
% 2.098 & 0.557 & 3.575 & 3.228
7 & 4.266 & 2.314 & 0.00028 & 0.52289 & 0.00311 \\
% 2.102 & 0.603 & 3.532 & 3.173
8 & 4.134 & 2.090 & 0.00036 & 0.30332 & 0.00201 \\
% 2.106 & 0.636 & 3.504 & 3.136
9 & 4.057 & 2.003 & 0.00039 & 0.84777 & 0.00108 \\
% 2.109 & 0.666 & 3.469 & 3.091
10 & 3.980 & 2.040 & 0.00033 & 2.86108 & 0.00052
 \\ \hline
16 & 15.28 & 6.508 & 0.00024 & 15.13 & 0.00047 \\
% 2.727 & 0.562 & 9.428 & 9.805
17 & 13.00 & 4.908 & 0.00037 & 9.12807 & 0.00085 \\
% 2.736 & 0.602 & 9.026 & 9.261
18 & 11.97 & 4.140 & 0.00047 & 7.78352 & 0.00116 \\
% 2.744 & 0.633 & 8.734 & 8.866
19 & 11.31 & 3.745 & 0.00053 & 4.9739 & 0.00203 \\
% 2.750 & 0.654 & 8.543 & 8.607
20 & 10.90 & 3.613 & 0.00053 & 3.90644 & 0.00266
 \\ \hline
27 & 18.56 & 6.396 & 0.00038 & 6.08298 & 0.00238 \\
% 2.871 & 0.609 & 11.25 & 12.13
28 & 15.49 & 4.938 & 0.00052 & 5.20018 & 0.00288 \\
% 2.882 & 0.647 & 10.64 & 11.25
29 & 14.13 & 4.374 & 0.00057 & 3.92533 & 0.00392 \\
% 2.884 & 0.658 & 10.42 & 10.94
30 & 13.69 & 4.374 & 0.00053 & 3.71457 & 0.00392
 \\ \hline
\end{tabular}
\end{ruledtabular}
 \protect\label{anal2}
%\end{scriptsize}
\end{minipage}
\end{table*}

%%%%%%%%%%%%%%%%%%%%%%%%%%%%%%%%%%%%%%%%%%%%%%%%%%%%%%%%%%%%%%%%%%%%%%%%%%%%%%%%%%%%%%5

\begin{table*}
 \centering
\begin{minipage}{0.5 \textwidth}
  \caption{EOS L.}
%  \begin{scriptsize}
  \centering
   \begin{ruledtabular}
\begin{tabular}{c c c c c c c c}
 $\#$ & $M$ & $j$ & $M_2$ & $S_3$ & $b$ & $\Delta M_2$(\%) & $\Delta S_3$(\%) \\
 \hline
1  & 2.080 & 0. & 0 & 0 & -0.25 & - & - \\ %& - & - & - & - & - & - \\
2  & 2.071 & 0.194 & -2.76 & -2.28 & -0.24 & 1.302 & 1.136 \\ %& -2.81 & -2.71 & 0.00001 & 0.00002 & 0.00028 & 0.00034 \\
3  & 2.075 & 0.324 & -7.55 & -10.5 & -0.24 & 1.328 & 1.156 \\ %& -15.6 & -15.4 & 0.00006 & 0.00002 & 0.00312 & 0.00328 \\
4  & 2.080 & 0.417 & -12.2 & -22.0 & -0.23 & 1.352 & 1.175 \\ %& -30.2 & -29.7 & 0.00015 & 0.00007 & 0.00445 & 0.00473 \\
5  & 2.083 & 0.483 & -16.2 & -33.9 & -0.23 & 1.369 & 1.188 \\ %& -43.6 & -42.9 & 0.00024 & 0.00014 & 0.00463 & 0.00500 \\
6  & 2.087 & 0.543 & -20.3 & -47.9 & -0.22 & 1.387 & 1.201 \\ %& -58.2 & -57.3 & 0.00034 & 0.00022 & 0.00413 & 0.00459 \\
7  & 2.090 & 0.598 & -24.4 & -63.5 & -0.22 & 1.399 & 1.210 \\ %& -73.4 & -72.2 & 0.00044 & 0.00030 & 0.00307 & 0.00360 \\
8  & 2.095 & 0.650 & -28.6 & -81.3 & -0.21 & 1.412 & 1.218 \\ %& -89.8 & -88.5 & 0.00052 & 0.00038 & 0.00164 & 0.00221 \\
9  & 2.096 & 0.690 & -32.1 & -97.1 & -0.21 & 1.417 & 1.220 \\ %& -103. & -102. & 0.00055 & 0.00041 & 0.00062 & 0.00102 \\
10  & 2.097 & 0.698 & -32.9 & -100. & -0.21 & 1.420 & 1.222 \\ %&-107. & -105. & 0.00055 & 0.00041 & 0.00059 & 0.00082\\
\hline
11  & 3.995 & 0. & 0 & 0 & -0.25 & - & - \\ %& - & - & - & - & - & - \\
12  & 4.012 & 0.178 & -3.80 & -4.23 & -0.24 & 13.65 & 16.18 \\ %& - & - & - & - & - & - \\
13  & 4.029 & 0.280 & -9.84 & -17.5 & -0.23 & 12.72 & 14.74 \\ %& - & - & - & - & - & - \\
14  & 4.051 & 0.375 & -18.5 & -45.4 & -0.22 & 11.94 & 13.57 \\ %& - & - & - & - & - & - \\
15  & 4.074 & 0.458 & -29.0 & -88.4 & -0.21 & 11.28 & 12.59 \\ %& - & - & - & - & - & - \\
16  & 4.098 & 0.528 & -40.3 & -144. & -0.20 & 10.72 & 11.77 \\ %& -150. & -126. & 0.00261 & 0.00039 & 0.01757 & 0.00158 \\
17  & 4.120 & 0.588 & -51.8 & -210. & -0.19 & 10.25 & 11.09 \\ %& -219. & -188. & 0.00307 & 0.00060 & 0.02042 & 0.00221 \\
18  & 4.139 & 0.635 & -62.6 & -279. & -0.19 & 9.865 & 10.54 \\ %& -288. & -250. & 0.00342 & 0.00083 & 0.02517 & 0.00347 \\
19  & 4.160 & 0.682 & -74.9 & -365. & -0.18 & 9.478 & 9.998 \\ %& -371. & -326. & 0.00360 & 0.00102 & 0.03060 & 0.00792 \\
20  & 4.167 & 0.700 & -79.8 & -401. & -0.17 & 9.335 & 9.794 \\ %&-405. & -357. & 0.00355 & 0.00104 & 0.03185 & 0.00960\\
\hline
21  & 4.321 & 0.478 & -29.0 & -88.0 & -0.20 & 17.68 & 22.66 \\ %& - & - & - & - & - & - \\
22  & 4.321 & 0.479 & -29.5 & -90.2 & -0.20 & 17.19 & 21.82 \\ %& - & - & - & - & - & - \\
23  & 4.324 & 0.483 & -30.7 & -95.7 & -0.20 & 16.66 & 20.90 \\ %& - & - & - & - & - & - \\
24  & 4.325 & 0.489 & -31.9 & -101. & -0.20 & 16.24 & 20.20 \\ %& - & - & - & - & - & - \\
25  & 4.333 & 0.505 & -35.0 & -116. & -0.20 & 15.55 & 19.03 \\ %& - & - & - & - & - & - \\
26  & 4.355 & 0.555 & -45.2 & -170. & -0.19 & 14.17 & 16.79 \\ %& -168. & -129. & 0.00402 & 0.00054 & 0.04037 & 0.00682 \\
27  & 4.377 & 0.602 & -55.9 & -233. & -0.19 & 13.27 & 15.36 \\ %& -235. & -190. & 0.00434 & 0.00071 & 0.03989 & 0.00600 \\
28  & 4.396 & 0.641 & -66.0 & -299. & -0.18 & 12.60 & 14.32 \\ %& -303. & -250. & 0.00459 & 0.00090 & 0.04279 & 0.00802 \\
29  & 4.418 & 0.684 & -78.7 & -389. & -0.17 & 11.92 & 13.28 \\ %& -391. & -328. & 0.00466 & 0.00108 & 0.04641 & 0.01207 \\
30  & 4.420 & 0.686 & -79.4 & -394. & -0.17 & 11.89 & 13.24 \\ %&-395. & -333. & 0.00466 & 0.00108 & 0.04653 & 0.01227
\end{tabular}
\end{ruledtabular}
%\end{scriptsize}
\protect\label{num3}
\end{minipage}
\end{table*}

\begin{table*}
 \centering
\begin{minipage}{0.5 \textwidth}
  \caption{EOS L.}
 % \begin{scriptsize}
  \centering
 \begin{ruledtabular}
\begin{tabular}{ c c c c c c}
% $M$ & $j$ & $\Delta Q$(\%) & $\Delta S_3$(\%)
$\#$  & $\Delta S_3^M$(\%) &
 $\frac{\sigma_{tt}|_{old}}{\sigma_{tt}|_{new}}$ & $\sigma_{tt}|_{new}$ &
 $\frac{\sigma_{t\phi}|_{old}}{\sigma_{t\phi}|_{new}}$ & $\sigma_{t\phi}|_{new}$
 \\ \hline
% 2.080 & 0. & - & - & - & 1 & - & 1 & - \\
% 2.071 & 0.194 & 1.302 & 1.136
2 & 3.364 & 0.846 & 0.00002 & 0.81731 & 0.00034 \\
% 2.075 & 0.324 & 1.328 & 1.156
3 & 1.677 & 2.888 & 0.00002 & 0.94960 & 0.00328 \\
% 2.080 & 0.417 & 1.352 & 1.175
4 & 1.579 & 2.017 & 0.00007 & 0.94067 & 0.00473 \\
% 2.083 & 0.483 & 1.369 & 1.188
5 & 1.550 & 1.679 & 0.00014 & 0.92563 & 0.00500 \\
% 2.087 & 0.543 & 1.387 & 1.201
6 & 1.540 & 1.519 & 0.00022 & 0.90005 & 0.00459 \\
% 2.090 & 0.598 & 1.399 & 1.210
7 & 1.534 & 1.431 & 0.00030 & 0.85334 & 0.00360 \\
% 2.095 & 0.650 & 1.412 & 1.218
8 & 1.535 & 1.377 & 0.00038 & 0.74334 & 0.00221 \\
% 2.096 & 0.690 & 1.417 & 1.220
9 & 1.533 & 1.355 & 0.00041 & 0.61091 & 0.00102 \\
% 2.097 & 0.698 & 1.420 & 1.222
10 & 1.534 & 1.354 & 0.00041 & 0.72087 & 0.00082
 \\ \hline
16 & 16.01 & 6.594 & 0.00039 & 11.1166 & 0.00158 \\
% 4.120 & 0.588 & 10.25 & 11.09
17 & 14.15 & 5.081 & 0.00060 & 9.2408 & 0.00221 \\
% 4.139 & 0.635 & 9.865 & 10.54
18 & 13.08 & 4.119 & 0.00083 & 7.25373 & 0.00347 \\
% 4.160 & 0.682 & 9.478 & 9.998
19 & 12.18 & 3.520 & 0.00102 & 3.86123 & 0.00792 \\
% 4.167 & 0.700 & 9.335 & 9.794
20 & 11.87 & 3.397 & 0.00104 & 3.31502 & 0.00960
 \\ \hline
26 & 23.13 & 7.416 & 0.00054 & 5.91821 & 0.00682 \\
% 4.377 & 0.602 & 13.27 & 15.36
27 & 19.25 & 6.100 & 0.00071 & 6.64409 & 0.00600 \\
% 4.396 & 0.641 & 12.60 & 14.32
28 & 17.44 & 5.094 & 0.00090 & 5.33573 & 0.00802 \\
% 4.418 & 0.684 & 11.92 & 13.28
29 & 15.89 & 4.316 & 0.00108 & 3.84442 & 0.01207 \\
% 4.420 & 0.686 & 11.89 & 13.24
30 & 15.82 & 4.286 & 0.00108 & 3.79143 & 0.01227

\end{tabular}
\end{ruledtabular}
%\end{scriptsize}
 \protect\label{anal3}
\end{minipage}
\end{table*}

%%%%%%%%%%%%%%%%%%%%%%%%%%%%%%%%%%%%%%%%%%%%%%%%%%%%%%%%%%%%%%%%%%%%%%%%%%%%%%%%%%%%%%5

%%%%%%%%%%%%%%%%%%%%%%%%%%%%%%%%%%%%%%%%%%%%%%%
\subsection{Results for the $R_{\textrm{ISCO}}$}
\label{sec:isco}
%%%%%%%%%%%%%%%%%%%%%%%%%%%%%%%%%%%%%%%%%%%%%%%

For the assessment of the implications the correction in the
multipole moments has, we have also calculated its effect on the
location of the radius of the innermost stable circular orbit
$R_{ISCO}$. The ISCO of the co-rotating orbits for the numerical
neutron star models is directly computed by the ``rns'' code. For
every model we have computed the ISCOs for the analytic metric of
Manko et al. using the wrong quadrupole and the corrected one. The
same is repeated using the formula of Shibata and Sasaki
\cite{ShibSasa} up to terms corresponding to the octupole
current-mass moment (as it was done in \cite{BertSter}).

Tables \ref{iscoTab1}, \ref{iscoTab2}, \ref{iscoTab3}, show our
results for the $R_{ISCO}$, namely the numerical ISCO, the Manko
et al. ISCO, before and after the correction in the quadrupole,
and the Shibata and Sasaki ISCO, again before and after the
correction. The models for which there is no value for the
numerical $R_{ISCO}$, correspond to neutron stars that have their
ISCO below the star's surface. The models that do not have
analytic $R_{ISCO}$ correspond to cases for which no analytic
metric can be found to match their numerical quadrupole.

The $R_{ISCO}$ calculated from the Manko et al. metric is
relatively close to the numerical one before the correction and it
gets even closer after the correction. As discussed in the text,
the ISCO calculated from the Shibata and Sasaki formula is already
a good approximation, before the correction, of the numerical one
for slow rotation, and improves slightly after the correction. On
the other hand, for higher rotation the improvement is greater but
then the approximate formula is insufficient to describe so
rapidly rotating stars, at least up to the approximation used.

\begin{table*}
 \centering
\begin{minipage}{0.5 \textwidth}
  \caption{$R_{ISCO}$ for the EOS AU.}
 % \begin{scriptsize}
  \centering
 \begin{ruledtabular}
\begin{tabular}{ c c c c c c}
% $M$ & $j$ & $\Delta Q$(\%) & $\Delta S_3$(\%)
$\#$  & $R_{ISCO}^n$ & $R_{ISCO}^M|_{old}$ & $R_{ISCO}^M|_{new}$&
 $R_{ISCO}^{SS}|_{old}$ & $R_{ISCO}^{SS}|_{new}$
 \\ \hline
 1 & 12.41 & - & - & 12.41 & 12.41 \\
 2 & 11.36 & - & - & 11.30 & 11.32 \\
 3 & 11.11 & 11.02 & 11.05 & 10.89 & 10.93 \\
 4 & - & 10.84 & 10.89 & 10.63 & 10.69 \\
 5 & - & 10.80 & 10.87 & 10.50 & 10.59 \\
 6 & - & 10.82 & 10.91 & 10.41 & 10.52 \\
 7 & - & 10.88 & 10.98 & 10.35 & 10.48 \\
 8 & - & 10.94 & 11.06 & 10.31 & 10.47 \\
 9 & - & 11.02 & 11.14 & 10.28 & 10.45 \\
 10 & - & 11.05 & 11.17 & 10.27 & 10.45
 \\ \hline
 11 & 18.91 & - & - & 18.91 & 18.91 \\
 12 & 17.00 & - & - & 16.94 & 16.97 \\
 13 & 15.97 & - & - & 15.76 & 15.83 \\
 14 & 15.21 & - & - & 14.73 & 14.86 \\
 15 & 14.68 & - & - & 13.86 & 14.03 \\
 16 & 14.35 & - & - & 13.10 & 13.31 \\
 17 & 14.16 & - & - & 12.44 & 12.68 \\
 18 & 14.08 & 13.42 & 13.91 & 11.88 & 12.15 \\
 19 & 14.08 & 13.29 & 13.83 & 11.44 & 11.74 \\
 20 & 14.12 & 13.20 & 13.79 & 11.01 & 11.32
 \\ \hline
 21 & 14.80 & - & - & 13.99 & 14.19 \\
 22 & 14.81 & - & - & 13.99 & 14.19 \\
 23 & 14.80 & - & - & 13.95 & 14.16 \\
 24 & 14.77 & - & - & 13.88 & 14.09 \\
 25 & 14.64 & - & - & 13.55 & 13.77 \\
 26 & 14.45 & - & - & 13.03 & 13.28 \\
 27 & 14.31 & - & - & 12.49 & 12.76 \\
 28 & 14.24 & - & - & 12.02 & 12.31 \\
 29 & 14.23 & 13.36 & 14.04 & 11.48 & 11.79 \\
 30 & 14.25 & 13.28 & 14.01 & 11.18 & 11.50

\end{tabular}
\end{ruledtabular}
%\end{scriptsize}
\protect\label{iscoTab1}
\end{minipage}
\end{table*}

\begin{table*}
 \centering
\begin{minipage}{0.5 \textwidth}
  \caption{$R_{ISCO}$ for the EOS FPS.}
 % \begin{scriptsize}
  \centering
 \begin{ruledtabular}
\begin{tabular}{ c c c c c c}
% $M$ & $j$ & $\Delta Q$(\%) & $\Delta S_3$(\%)
$\#$  & $R_{ISCO}^n$ & $R_{ISCO}^M|_{old}$ & $R_{ISCO}^M|_{new}$&
 $R_{ISCO}^{SS}|_{old}$ & $R_{ISCO}^{SS}|_{new}$
 \\ \hline
 1 & 12.40 & - & - & 12.40 & 12.40 \\
 2 & 11.38 & - & - & 11.31 & 11.33 \\
 3 & - & 11.09 & 11.12 & 10.98 & 11.01 \\
 4 & - & 10.96 & 11.00 & 10.78 & 10.83 \\
 5 & - & 10.93 & 10.99 & 10.67 & 10.75 \\
 6 & - & 10.98 & 11.06 & 10.62 & 10.72 \\
 7 & - & 11.06 & 11.15 & 10.60 & 10.72 \\
 8 & - & 11.18 & 11.28 & 10.60 & 10.74 \\
 9 & - & 11.28 & 11.39 & 10.61 & 10.77 \\
 10 & - & 11.40 & 11.51 & 10.63 & 10.80
 \\ \hline
 11 & 15.94 & - & - & 15.94 & 15.94 \\
 12 & 14.61 & - & - & 14.58 & 14.59 \\
 13 & 13.94 & - & - & 13.81 & 13.86 \\
 14 & 13.44 & - & - & 13.13 & 13.20 \\
 15 & 13.11 & - & - & 12.47 & 12.59 \\
 16 & 13.00 & 12.70 & 12.87 & 11.99 & 12.14 \\
 17 & 13.04 & 12.55 & 12.77 & 11.52 & 11.71 \\
 18 & 13.14 & 12.52 & 12.76 & 11.21 & 11.42 \\
 19 & 13.27 & 12.53 & 12.78 & 10.97 & 11.19 \\
 20 & - & 12.55 & 12.82 & 10.80 & 11.04
 \\ \hline
 21 & 13.32 & - & - & 12.83 & 12.96 \\
 22 & 13.33 & - & - & 12.84 & 12.97 \\
 23 & 13.34 & - & - & 12.81 & 12.94 \\
 24 & 13.33 & - & - & 12.77 & 12.90 \\
 25 & 13.30 & - & - & 12.69 & 12.83 \\
 26 & 13.21 & - & - & 12.34 & 12.50 \\
 27 & 13.17 & 12.78 & 13.02 & 11.84 & 12.03 \\
 28 & 13.27 & 12.68 & 12.98 & 11.33 & 11.56 \\
 29 & 13.43 & 12.68 & 13.00 & 10.99 & 11.24 \\
 30 & - & 12.69 & 13.02 & 10.87 & 11.13

\end{tabular}
\end{ruledtabular}
%\end{scriptsize}
\protect\label{iscoTab2}
\end{minipage}
\end{table*}

\begin{table*}
 \centering
\begin{minipage}{0.5 \textwidth}
  \caption{$R_{ISCO}$ for the EOS L.}
 % \begin{scriptsize}
  \centering
 \begin{ruledtabular}
\begin{tabular}{ c c c c c c}
% $M$ & $j$ & $\Delta Q$(\%) & $\Delta S_3$(\%)
$\#$  & $R_{ISCO}^n$ & $R_{ISCO}^M|_{old}$ & $R_{ISCO}^M|_{new}$&
 $R_{ISCO}^{SS}|_{old}$ & $R_{ISCO}^{SS}|_{new}$
 \\ \hline
 1 & - & - & - & 12.48 & 12.48 \\
 2 & - & 11.73 & 11.74 & 11.69 & 11.70 \\
 3 & - & 11.80 & 11.82 & 11.84 & 11.87 \\
 4 & - & 12.09 & 12.12 & 12.20 & 12.25 \\
 5 & - & 12.38 & 12.41 & 12.57 & 12.63 \\
 6 & - & 12.69 & 12.74 & 12.98 & 13.06 \\
 7 & - & 13.01 & 13.06 & 13.40 & 13.50 \\
 8 & - & 13.34 & 13.40 & 13.85 & 13.97 \\
 9 & - & 13.61 & 13.67 & 14.22 & 14.36 \\
 10 & - & 13.67 & 13.73 & 14.30 & 14.44
 \\ \hline
 11 & 23.97 & - & - & 23.97 & 23.97 \\
 12 & 21.83 & - & - & 21.74 & 21.77 \\
 13 & 20.78 & - & - & 20.45 & 20.51 \\
 14 & 20.00 & - & - & 19.17 & 19.27 \\
 15 & 19.52 & - & - & 17.96 & 18.09 \\
 16 & 19.32 & 18.80 & 19.1 & 16.81 & 16.96 \\
 17 & 19.32 & 18.56 & 18.92 & 15.71 & 15.89 \\
 18 & 19.46 & 18.46 & 18.88 & 14.73 & 14.92 \\
 19 & 19.73 & 18.46 & 18.94 & 13.63 & 13.82 \\
 20 & 19.87 & 18.49 & 18.98 & 13.19 & 13.39
 \\ \hline
 21 & 19.81 & - & - & 18.37 & 18.52 \\
 22 & 19.84 & - & - & 18.37 & 18.52 \\
 23 & 19.85 & - & - & 18.30 & 18.45 \\
 24 & 19.85 & - & - & 18.21 & 18.37 \\
 25 & 19.80 & - & - & 17.94 & 18.10 \\
 26 & 19.68 & 19.15 & 19.54 & 16.99 & 17.16 \\
 27 & 19.67 & 18.90 & 19.39 & 16.03 & 16.21 \\
 28 & 19.76 & 18.78 & 19.33 & 15.13 & 15.31 \\
 29 & 19.98 & 18.74 & 19.35 & 14.02 & 14.20 \\
 30 & 19.99 & 18.74 & 19.35 & 13.96 & 14.14

\end{tabular}
\end{ruledtabular}
%\end{scriptsize}
\protect\label{iscoTab3}
\end{minipage}
\end{table*}

\end{document}